\pdfoutput=1

\documentclass[prb,aps,twocolumn]{revtex4}

\usepackage{bm}
\usepackage{amsmath}
\usepackage{amssymb}

\usepackage{graphicx}
\begin{document}



\title{Superconducting pairing symmetry on the extended Hubbard model in the presence of the Rashba-type spin-orbit coupling}


\author{Keisuke Shigeta}
\author{Seiichiro Onari}
\author{Yukio Tanaka}

\affiliation{Department of Applied Physics, 
Nagoya University, Nagoya 464-8603, Japan}

\begin{abstract}
In order to study the pairing symmetry in non-centrosymmetric 
superconductors, we solve the linearized ${\acute{\mathrm{E}}}$liashberg's 
equation on the two-dimensional extended Hubbard model 
in the presence of the Rashba-type 
spin-orbit coupling (RSOC) within the random phase approximation. 
In the presence of the RSOC, three types of pairing symmetries 
appear in the phase diagram with respect to 
the on-site Coulomb repulsion $U$ and off-site one $V$. 
Each of pairing symmetries is admixture of spin-singlet and -triplet ones. 
On the basis of analytical study, it is found that 
the admixture of spin-singlet and -triplet components 
depends on not only the predominant pairing 
symmetry but also dispersion relation and pairing interaction. 
\end{abstract}

\maketitle
%
%
\section{INTRODUCTION}
Since the discovery of superconductors without the inversion symmetry 
in CePt$_3$Si, \cite{CePt3Si} non-centrosymmetric superconductors 
has been studied intensively as unconventional superconductors. 
In particular, theoretical studies suggest interesting properties of 
non-centrosymmetric superconductors such as the magnetoelectric effect, 
\cite{magnetoelectric,magnetoelectric2,magnetoelectric3,magnetoelectric4,magnetoelectric5} 
anisotropic spin susceptibility, 
\cite{magnetoelectric5,chi,chi2,chi3,chi4,chi5} 
and the helical vortex state in magnetic fields. \cite{vortex,vortex2}
Today, there are various relevant systems in 
two-dimensional electron gas at heterointerface, $e.g.$ between SrTiO$_3$ 
and LaAlO$_3$, \cite{hetero} and non-centrosymmetric crystals, 
$e.g.$ CePt$_3$Si, \cite{CePt3Si} UIr, \cite{UIr} 
CeRhSi$_3$, \cite{CeRhSi3} CeIrSi$_3$, \cite{CeIrSi3} 
CeCoGe$_3$, \cite{CeCoGe3,CeCoGe3-2} and Li$_2$Pd$_x$Pt$_{3-x}$B. 
\cite{Li2Pt3B,Li2Pt3B-2,Li2Pt3B-3} 
\par
In superconductors with the inversion symmetry, pairing symmetry is classified 
into even- and odd-parity, $i.e.$ spin-singlet and -triplet. 
In the non-centrosymmetric superconductors, on the other hand, admixture 
of spin-singlet and -triplet pairings is realized. 
\cite{magnetoelectric,chi,chi2,admixture,admixture2} 
The admixture is induced by the antisymmetric spin-orbit coupling, which 
is generated by the lack of the inversion symmetry. 
For example, it has been proposed that admixture of spin-singlet $s$-wave 
and spin-triplet $p$-wave pairings is realized in a non-centrosymmetric 
heavy fermion superconductor CePt$_3$Si from both 
theoretical \cite{chi5,CePt3Si_theo,CePt3Si_theo2} and experimental 
\cite{CePt3Si,CePt3Si_exp,CePt3Si_exp2,CePt3Si_exp3,CePt3Si_exp4,CePt3Si_exp5,CePt3Si_exp6} 
studies. 
\par
In particular, non-centrosymmetric heavy fermion superconductors, $e.g.$ 
CePt$_3$Si, \cite{CePt3Si} UIr, \cite{UIr} CeRhSi$_3$, \cite{CeRhSi3} 
CeIrSi$_3$, \cite{CeIrSi3} and CeCoGe$_3$, \cite{CeCoGe3,CeCoGe3-2} 
are of interest 
because the superconductivity originates from the Coulomb repulsion. 
However, there are few theoretical studies on these materials on the basis 
of the microscopic calculation. 
\cite{chi5,CePt3Si_theo,NCS_HF_micro,NCS_HF_micro2} 
It is desired to microscopically understand the admixture of 
spin-singlet and -triplet pairings in non-centrosymmetric superconductors 
where Coulomb repulsion mediates pairing in more detail. 
\par
In order to study the above admixture, it is valuable to employ 
the extended Hubbard model because this model exhibits various 
pairing symmetries in the presence of the inversion symmetry. 
\cite{Onari,ext_Hub,ext_Hub2,ext_Hub2_1,ext_Hub3,ext_Hub4,ext_Hub5,ext_Hub6,ext_Hub7,ext_Hub8,ext_Hub9,ext_Hub10,ext_Hub11}
On the extended Hubbard model, the off-site Coulomb repulsion is considered 
in addition to the on-site one. 
It is well known that, while the on-site Coulomb repulsion induces 
the spin fluctuation, the charge fluctuation coexists with 
the spin fluctuation by introducing the off-site Coulomb repulsion. 
\cite{Onari,ext_Hub,ext_Hub2,ext_Hub2_1,ext_Hub3,ext_Hub4,ext_Hub5,ext_Hub6,ext_Hub7,ext_Hub8,ext_Hub9,ext_Hub10,ext_Hub11}
Due to the coexistence of the charge fluctuation with the spin one, 
especially on a two-dimensional square lattice near half-filling, 
three types of pairing symmetries, $i.e.$ spin-singlet $d_{x^2-y^2}$-wave, 
spin-triplet $f$-wave, and spin-singlet $d_{xy}$-wave ones, compete against 
each others. \cite{Onari}
\par
In the present study, in order to clarify pairing symmetry in 
non-centrosymmetric superconductors where the Coulomb repulsion mediates 
pairing, we investigate the two-dimensional extended Hubbard model 
in the presence of 
the Rashba-type spin-orbit coupling (RSOC) \cite{Rashba} on the basis of 
the random phase approximation (RPA). 
The RSOC induces breakdown of the inversion symmetry and 
admixture of pairing symmetry. 
\par
The admixture of pairing symmetry has already been studied on 
the extended Hubbard model in the presence of the RSOC on the basis of 
the RPA by Yokoyama $et$ $al$. \cite{NCS_HF_micro2} 
However, they investigated only the region where the off-site Coulomb 
repulsion is small, $i.e.$ spin-singlet $d_{x^2-y^2}$-wave pairing state 
is stable in the absence of the RSOC. 
Moreover, there were two simplifications in the pairing interaction. 
One is that they neglect cross terms of the bubble- and ladder-type diagrams, 
which are generated by the off-site Coulomb repulsion and the RSOC. 
The other is that the ladder-type diagrams with the off-site Coulomb 
repulsion are excluded. 
In the present study, there is no simplification described above. 
\par
The present paper is organized as follows. 
In \S\ref{sec_formulation}, we formulate the linearized 
${\acute{\mathrm{E}}}$liashberg's equation on the extended Hubbard model 
in the presence of the RSOC on the basis of the RPA. 
In \S\ref{sec_RPA}, we show results obtained by 
the numerical calculation within the RPA. 
After that, we discuss the pairing symmetry on the basis of analytical study 
in \S\ref{sec_analytical}. 
The summary is given in \S\ref{sec_summary}. 
%
%
\section{FORMULATION}
\label{sec_formulation}
We start with the two-dimensional extended Hubbard model 
in the presence of the RSOC. 
The Hamiltonian is given by 
\begin{align}
{\cal{H}}=
&\sum_{{\boldsymbol{k}},s}\varepsilon_{\boldsymbol{k}}c_{{\boldsymbol{k}}s}^{\dagger}c_{{\boldsymbol{k}}s}
+U\sum_{i}n_{i\uparrow}n_{i\downarrow}
+V\sum_{\langle i,j\rangle,s,s'}n_{is}n_{js'} \notag \\
&-\lambda\sum_{{\boldsymbol{k}},s,s'}\left[{\boldsymbol{g}}({\boldsymbol{k}})\cdot\hat{\boldsymbol{\sigma}}\right]_{ss'}c_{{\boldsymbol{k}}s}^{\dagger}c_{{\boldsymbol{k}}s'}, \\ 
\varepsilon_{\boldsymbol{k}}=&-2t(\cos k_x+\cos k_y)-4t'\cos k_x\cos k_y-\mu,
\end{align}
where $c_{{\boldsymbol{k}}s}^{(\dagger)}$ is an annihilation (a creation) 
operator for an electron with spin $s$ and momentum ${\boldsymbol{k}}$, 
$n_{is}$ is a number operator for an electron with spin $s$ at site $i$, 
and $\langle i,j\rangle$ denotes a set of the nearest neighbor sites. 
$\varepsilon_{\boldsymbol{k}}$ is the dispersion relation, 
where $t^{(\prime)}$ is the (second) nearest neighbor hopping on a square 
lattice and $\mu$ is the chemical potential. 
We consider the on-site Coulomb repulsion $U$ and the off-site one $V$ 
between the nearest neighbor sites. 
The fourth term is the RSOC, where $\lambda$ is a magnitude of the RSOC 
and $\hat{\boldsymbol{\sigma}}$ are the Pauli matrices. 
The vector ${\boldsymbol{g}}({\boldsymbol{k}})$ with the relation 
${\boldsymbol{g}}({\boldsymbol{k}})=-{\boldsymbol{g}}(-{\boldsymbol{k}})$ 
induces breakdown of the inversion symmetry. 
We adopt ${\boldsymbol{g}}({\boldsymbol{k}})=(-v_y({\boldsymbol{k}}),v_x({\boldsymbol{k}}),0)/\bar{v}$ 
with the quasiparticle velocity 
$v_{x(y)}({\boldsymbol{k}})=\partial\varepsilon_{\boldsymbol{k}}/\partial k_{x(y)}=2t\sin k_{x(y)}+4t'\sin k_{x(y)}\cos k_{y(x)}$. 
${\boldsymbol{g}}({\boldsymbol{k}})$ is normalized by the average velocity 
$\bar{v}$ which is given by 
$\bar{v}^2=\sum_{\boldsymbol{k}}\left[v_x({\boldsymbol{k}})^2+v_y({\boldsymbol{k}})^2\right]/N$,
where $N$ is the number of ${\boldsymbol{k}}$-meshes. 
The bare Green's function is given by the following $2\times2$ matrix 
in spin space, 
\begin{align}
\hat{G}(k)=&\left(
\begin{array}{cc}
G_{\uparrow\uparrow}(k) & G_{\uparrow\downarrow}(k) \\
G_{\downarrow\uparrow}(k) & G_{\downarrow\downarrow}(k) \\
\end{array}
\right) \notag \\ 
=&\left[
({\mathrm{i}}\omega_n-\varepsilon_{\boldsymbol{k}})\hat{\mathrm{I}}
+\lambda{\boldsymbol{g}}({\boldsymbol{k}})\cdot\hat{\boldsymbol{\sigma}}
\right]^{-1},
\end{align}
where $\hat{\mathrm{I}}$ is a unit matrix and 
$k\equiv({\mathrm{i}}\omega_n,{\boldsymbol{k}})$ is an abbreviation. 
$\omega_n=(2n-1)\pi T$ is the Matsubara frequency for fermions, 
where $n$ is an integer and $T$ is temperature. 
\par
In order to estimate the pairing instability, we solve 
the linearized ${\acute{\mathrm{E}}}$liashberg's equation 
within the RPA 
\begin{align}
\alpha\Delta_{s_1s_2}(k)=&-\frac{T}{N}\sum_{k',c_1,c_2,s_3,s_4}\Gamma_{s_1s_2s_3s_4}^{c_1c_2}(k-k') \notag \\
&{\hspace{15mm}}\times P^{c_1}({\boldsymbol{k'}})P^{c_2}(-{\boldsymbol{k}})F_{s_3s_4}(k'), \label{eq_eliash1} \\
F_{s_1s_2}(k)=&\sum_{s_3,s_4}G_{s_1s_3}(k)G_{s_2s_4}(-k)\Delta_{s_3s_4}(k),
\label{eq_eliash2}
\end{align}
where 
${\boldsymbol{P}}({\boldsymbol{k}})=(1,\cos k_x,\sin k_x,\cos k_y,\sin k_y)$ 
is the phase factor which originates from ladder-type connections with 
the off-site Coulomb repulsion $V({\boldsymbol{q}})=2V(\cos q_x+\cos q_y)$ 
in the diagrammatic expression. 
The linearized ${\acute{\mathrm{E}}}$liashberg's equation 
(\ref{eq_eliash1}) and (\ref{eq_eliash2}) is an eigenvalue equation 
whose eigenvalue and eigenfunction are $\alpha$ and $\hat{\Delta}(k)$, 
respectively. 
When the eigenvalue $\alpha$ reaches unity, temperature $T$ corresponds to 
the superconducting transition temperature $T_{\mathrm{C}}$. 
Thus, the eigenvalue $\alpha$ implies the pairing instability 
with the gap function $\hat{\Delta}(k)$. 
In solving the linearized ${\acute{\mathrm{E}}}$liashberg's 
equation, we employ the implicit restarted Arnoldi method. \cite{Arnoldi} 
This method is powerful in solving an eigenvalue equation with 
nearly degenerate solutions. 
\par
The spin-singlet and -triplet components with $S_z=0$ are extracted by 
$\left[\Delta_{\uparrow\downarrow}(k)\pm\Delta_{\downarrow\uparrow}(k)\right]/2$, 
where the sign $+(-)$ corresponds to spin-triplet (-singlet) one. 
The spin-triplet components with $S_z=\pm1$ are given by 
$\Delta_{\uparrow\uparrow(\downarrow\downarrow)}(k)$ for the sign $+(-)$. 
In the present paper, we choose the solution whose spin-singlet component 
is real. 
The spin-triplet ($S_z=\pm1$) components are imaginary. 
The real and imaginary parts of the spin-triplet ($S_z=\pm1$) components 
have same amplitude while nodes of the real and imaginary parts have 
the relation of rotation around ${\boldsymbol{k}}=(0,0)$. 
\par
\begin{figure}[htbp]
\includegraphics[width=0.99\linewidth,keepaspectratio]
                  {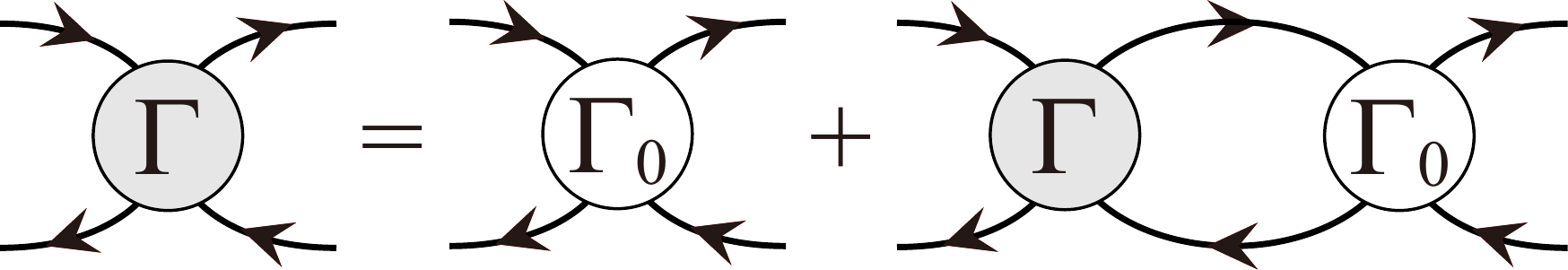}
 \caption{Diagrammatic sense of the effective pairing interaction $\hat{\Gamma}(q)$ within the RPA. $\hat{\Gamma}_0(q)$ includes the on- and off-site Coulomb repulsions.}
\label{fig_RPA}
\end{figure}

Within the RPA, the effective pairing interaction 
$\Gamma_{s_1s_2s_3s_4}^{c_1c_2}(q)$, 
where $q\equiv({\mathrm{i}}\nu_m,{\boldsymbol{q}})$ with 
the Matsubara frequency for bosons $\nu_m=2m\pi T$, 
is obtained by collecting the infinite series which consist of 
the irreducible susceptibility in the diagrammatic expression. 
The irreducible susceptibility is given by 
\begin{align}
\chi_{0,s_1s_2s_3s_4}^{c_1c_2}(q)=&-\frac{T}{N}\sum_{k}G_{s_1s_3}(k+q)G_{s_4s_2}(k) \notag \\
&\hspace{25mm}\times P^{c_1}({\boldsymbol{k}})P^{c_2}({\boldsymbol{k}}).
\end{align}
The dressed susceptibility is given by 
\begin{align}
\hat{\chi}(q)=\hat{\chi}_0(q)\left[\hat{\mathrm{I}}-\hat{\Gamma}_0({\boldsymbol{q}})\hat{\chi}_0(q)\right]^{-1},
\label{eq_chi}
\end{align}
where the matrices are $20\times20$ ones with spin indices $s_i$ and 
phase factor ones $c_i$ defined as 
\begin{align}
\hat{M}=&\left(
\begin{array}{ccccc}
\hat{M}^{11} & \hat{M}^{12} & \hat{M}^{13} & \hat{M}^{14} & \hat{M}^{15} \\
\hat{M}^{21} & \hat{M}^{22} & \hat{M}^{23} & \hat{M}^{24} & \hat{M}^{25} \\
\hat{M}^{31} & \hat{M}^{32} & \hat{M}^{33} & \hat{M}^{34} & \hat{M}^{35} \\
\hat{M}^{41} & \hat{M}^{42} & \hat{M}^{43} & \hat{M}^{44} & \hat{M}^{45} \\
\hat{M}^{51} & \hat{M}^{52} & \hat{M}^{53} & \hat{M}^{54} & \hat{M}^{55} \\
\end{array}
\right), \\
\hat{M}^{c_1c_2}=&\left(
\begin{array}{cccc}
M_{  \uparrow  \uparrow  \uparrow  \uparrow}^{c_1c_2} &
M_{  \uparrow  \uparrow  \uparrow\downarrow}^{c_1c_2} &
M_{  \uparrow  \uparrow\downarrow  \uparrow}^{c_1c_2} &
M_{  \uparrow  \uparrow\downarrow\downarrow}^{c_1c_2} \\
M_{  \uparrow\downarrow  \uparrow  \uparrow}^{c_1c_2} &
M_{  \uparrow\downarrow  \uparrow\downarrow}^{c_1c_2} &
M_{  \uparrow\downarrow\downarrow  \uparrow}^{c_1c_2} &
M_{  \uparrow\downarrow\downarrow\downarrow}^{c_1c_2} \\
M_{\downarrow  \uparrow  \uparrow  \uparrow}^{c_1c_2} &
M_{\downarrow  \uparrow  \uparrow\downarrow}^{c_1c_2} &
M_{\downarrow  \uparrow\downarrow  \uparrow}^{c_1c_2} &
M_{\downarrow  \uparrow\downarrow\downarrow}^{c_1c_2} \\
M_{\downarrow\downarrow  \uparrow  \uparrow}^{c_1c_2} &
M_{\downarrow\downarrow  \uparrow\downarrow}^{c_1c_2} &
M_{\downarrow\downarrow\downarrow  \uparrow}^{c_1c_2} &
M_{\downarrow\downarrow\downarrow\downarrow}^{c_1c_2} \\
\end{array}
\right).
\end{align}
The matrix $\hat{\Gamma}_0({\boldsymbol{q}})$ is given by 
\begin{align}
\hat{\Gamma}_0^{11}({\boldsymbol{q}})=&\left(
\begin{array}{cccc}
  -V({\boldsymbol{q}}) & 0 & 0 & -U-V({\boldsymbol{q}}) \\
0                      & U & 0 & 0                      \\
0                      & 0 & U & 0                      \\
-U-V({\boldsymbol{q}}) & 0 & 0 &   -V({\boldsymbol{q}}) \\
\end{array}
\right), \\
\hat{\Gamma}_0^{c_1c_2}({\boldsymbol{q}})=&\left(
\begin{array}{cccc}
2V & 0  & 0  & 0  \\
0  & 2V & 0  & 0  \\
0  & 0  & 2V & 0  \\
0  & 0  & 0  & 2V \\
\end{array}
\right) \notag \\
&\hspace{30mm}(c_1=c_2=2-5), \\
\hat{\Gamma}_0^{c_1c_2}({\boldsymbol{q}})=&\hat{0}
\hspace{28mm}(c_1\neq c_2).
\end{align}
By using the dressed susceptibility, the effective pairing interaction is 
expressed as 
\begin{align}
\Gamma_{s_1s_2s_3s_4}^{c_1c_2}(q)=&
-\left[\hat{\Gamma}_0({\boldsymbol{q}})\hat{\chi}(q)\hat{\Gamma}_0({\boldsymbol{q}})\right]_{s_1s_3s_4s_2}^{c_1c_2} \notag \\
&-\Gamma_{0,s_1s_2s_3s_4}^{\prime c_1c_2}({\boldsymbol{q}}),
\end{align}
where the matrix $\hat{\Gamma}_0^{\prime}({\boldsymbol{q}})$ is given by 
\begin{align}
\hat{\Gamma}_0^{\prime11}({\boldsymbol{q}})=&\left(
\begin{array}{cccc}
-V({\boldsymbol{q}}) & 0 & 0 & 0 \\
0 & -U-V({\boldsymbol{q}}) & 0 & 0 \\
0 & 0 & -U-V({\boldsymbol{q}}) & 0 \\
0 & 0 & 0 & -V({\boldsymbol{q}}) \\
\end{array}
\right), \\
\hat{\Gamma}_0^{\prime c_1c_2}({\boldsymbol{q}})=&\hat{0}
\hspace{5mm}(c_1\neq1,c_2\neq1).
\end{align}
Fig. \ref{fig_RPA} shows a sense of the formulation for 
the pairing interaction within the RPA in a diagrammatic representation. 
In the present RPA for the extended Hubbard model, cross terms of 
the bubble- and ladder-type diagrams, which generated by the off-site Coulomb 
repulsion and the RSOC, are taken into account. 
\par
Within the RPA, the spin susceptibility is expressed as 
\begin{align}
\chi_{\mathrm{sp}}^{\xi\eta}(q)=&\frac{1}{N}\int_{0}^{\beta}{\mathrm{d}}\tau
{\mathrm{e}}^{{\mathrm{i}}\nu_m\tau}
\langle T_{\tau}S^{\xi}(\tau,{\boldsymbol{q}})
S^{\eta}(-{\boldsymbol{q}})\rangle \\
=&\frac{1}{4}\sum_{s_1,s_2,s_3,s_4}\sigma_{s_3s_4}^{\xi}
\chi_{s_1s_2s_3s_4}^{11}(q)\sigma_{s_2s_1}^{\eta},
\end{align}
where 
\begin{align}
S^{\xi}({\boldsymbol{q}})=&\frac{1}{2}\sum_{{\boldsymbol{k}},s,s'}
\sigma_{ss'}^{\xi}c_{{\boldsymbol{k}}+{\boldsymbol{q}}s}^{\dagger}
c_{{\boldsymbol{k}}s'}, \\
S^{\xi}(\tau,{\boldsymbol{q}})=&{\mathrm{e}}^{{\cal{H}}\tau}
S^{\xi}({\boldsymbol{q}}){\mathrm{e}}^{-{\cal{H}}\tau},
\end{align}
with $\xi,\eta=x,y,z$. 
Similarly, the charge susceptibility is expressed as 
\begin{align}
\chi_{\mathrm{ch}}(q)=&\frac{1}{2N}\int_{0}^{\beta}{\mathrm{d}}\tau
{\mathrm{e}}^{{\mathrm{i}}\nu_m\tau}
\langle T_{\tau}\rho(\tau,{\boldsymbol{q}})
\rho(-{\boldsymbol{q}})\rangle \\
=&\frac{1}{2}\sum_{s,s'}\chi_{sss's'}^{11}(q),
\end{align}
where 
\begin{align}
\rho({\boldsymbol{q}})=&\sum_{\boldsymbol{k}}
(c_{{\boldsymbol{k}}+{\boldsymbol{q}}\uparrow}^{\dagger}
c_{{\boldsymbol{k}}\uparrow}
+c_{{\boldsymbol{k}}+{\boldsymbol{q}}\downarrow}^{\dagger}
c_{{\boldsymbol{k}}\downarrow}), \\
\rho(\tau,{\boldsymbol{q}})=&{\mathrm{e}}^{{\cal{H}}\tau}
\rho({\boldsymbol{q}}){\mathrm{e}}^{-{\cal{H}}\tau}.
\end{align}
\par
In the present paper, we choose $t=1$ for a unit of energy. 
The second nearest neighbor hopping, temperature, and filling are always 
$t'=0.1$, $T=0.04$, and $0.8$ electrons per site, respectively, 
in the actual numerical calculation. 
We take $64\times64$ ${\boldsymbol{k}}$-meshes and 
$1024$ Matsubara frequencies. 
%
%
\section{RESULTS}
\subsection{Numerical calculation within the RPA}
\label{sec_RPA}
In this subsection, we show results obtained by the numerical calculation 
within the RPA. 
\subsubsection{In the absence of the RSOC}
\begin{figure}[htbp]
\includegraphics[width=0.99\linewidth,keepaspectratio]
                  {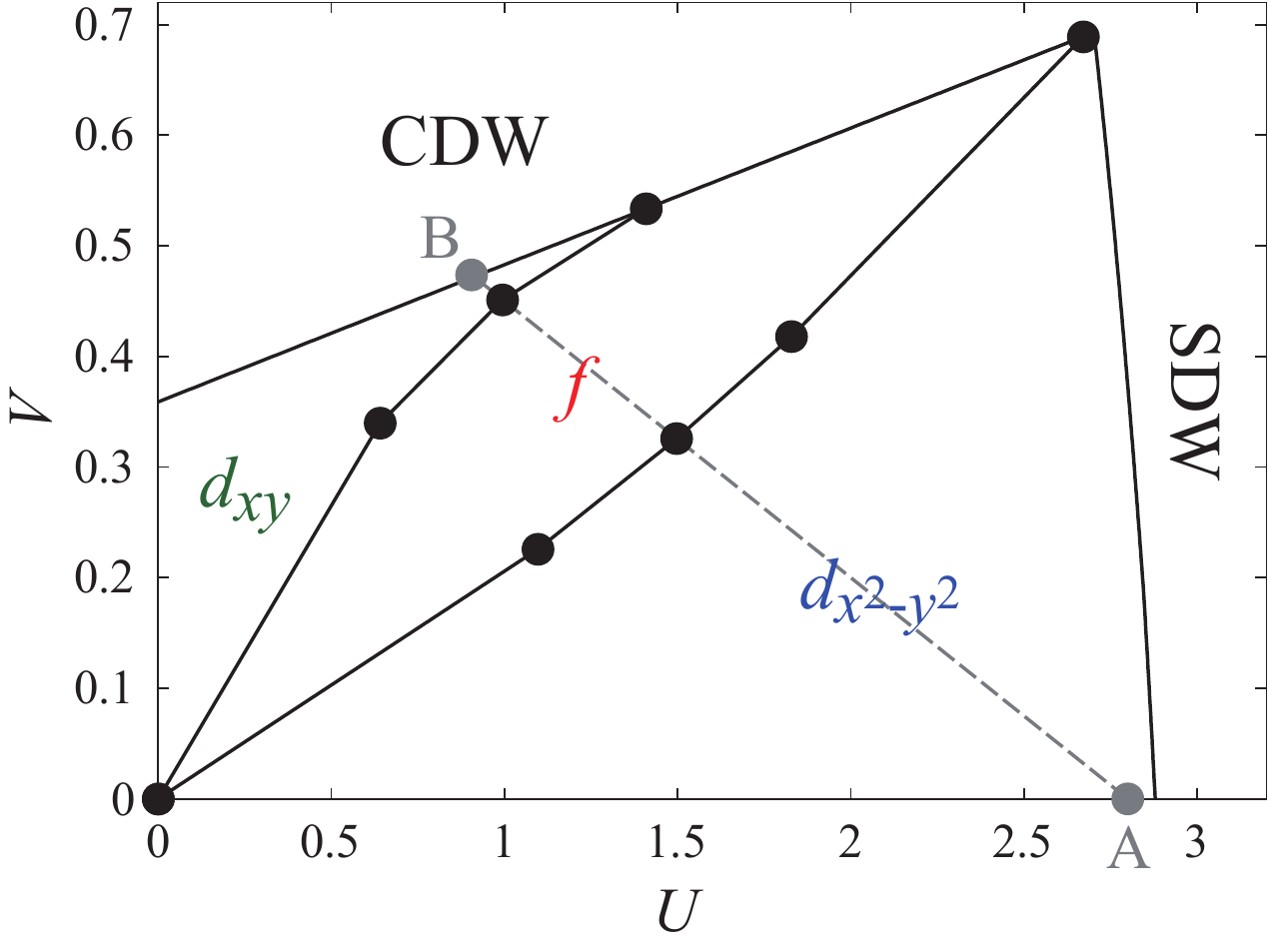}
 \caption{(Color online) $U$-$V$ phase diagram in the absence of the RSOC within the RPA. The broken line shows $V=-(U-2.8)/4$ between points A $(2.8,0)$ and B $(0.9,0.475)$.}
\label{fig_U-V-phase_lambda0}
\end{figure}

First, we check the pairing symmetry in the absence of the RSOC, 
$i.e.$ $\lambda=0$. 
Fig. \ref{fig_U-V-phase_lambda0} shows $U$-$V$ phase diagram. 
We identify the pairing symmetry with the largest eigenvalue $\alpha$ 
in the linearized ${\acute{\mathrm{E}}}$liashberg's equation 
(\ref{eq_eliash1}) and (\ref{eq_eliash2}) at $T=0.04$. 
The boundary with the spin- or charge-density-wave (SDW or CDW) phase is 
determined by the Stoner's factor, which is defined as the largest 
eigenvalue of the matrix $\hat{\Gamma}_0({\boldsymbol{q}})\hat{\chi}_0(q)$ 
in the dressed susceptibility (\ref{eq_chi}). 
When the Stoner's factor reaches unity, the dressed susceptibility diverges. 
In the present paper, we define the SDW or CDW phase as 
the region where the Stoner's factor reaches $0.98$. 
As shown in Fig. \ref{fig_U-V-phase_lambda0}, 
three types of the pairing symmetries can appear 
by tuning $U$ and $V$; 
spin-singlet $d_{x^2-y^2}$-wave, spin-triplet $f$-wave, 
and spin-singlet $d_{xy}$-wave pairing symmetries. 
\par
\begin{figure}[htbp]
\includegraphics[width=0.99\linewidth,keepaspectratio]
                  {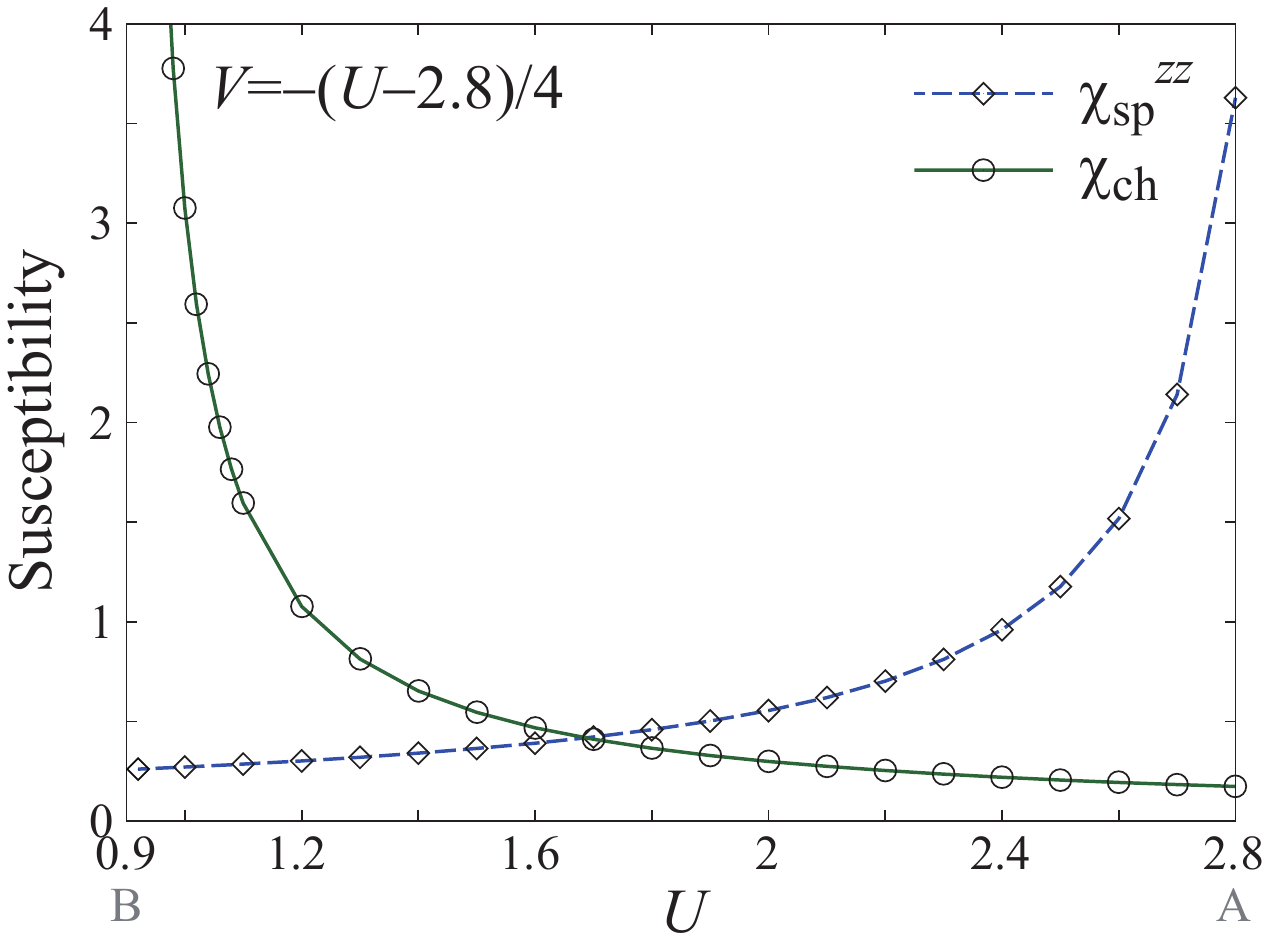}
 \caption{(Color online) The spin and charge susceptibilities on the broken line $V=-(U-2.8)/4$ in Fig. \ref{fig_U-V-phase_lambda0} ($U$-$V$ space) without the RSOC. The maximum values are plotted for each $(U,V)$-point. The result is numerically obtained within the RPA. Points A and B correspond to those in Fig. \ref{fig_U-V-phase_lambda0}.}
\label{fig_U_chi_lambda0}
\end{figure}

Three types of the pairing symmetries are caused by 
the cooperative/competitive spin and charge susceptibilities controlled 
by $U$ and $V$. \cite{Onari} 
Fig. \ref{fig_U_chi_lambda0} shows the spin and charge susceptibilities 
in the absence of the RSOC on the broken line $V=-(U-2.8)/4$ 
in Fig. \ref{fig_U-V-phase_lambda0} ($U$-$V$ space). 
Here, we plot the maximum values of $\chi_{\mathrm{sp}}^{zz}(q)$ and 
$\chi_{\mathrm{ch}}(q)$ for each $(U,V)$-point. 
Note that, in the absence of the RSOC, 
$\chi_{\mathrm{sp}}^{xx}(q)=\chi_{\mathrm{sp}}^{yy}(q)=\chi_{\mathrm{sp}}^{zz}(q)$ 
and $\chi_{\mathrm{sp}}^{\xi\eta}(q)=0$ for $\xi\neq\eta$, 
$i.e.$ the spin susceptibility is isotropic. 
In the region where $U$ ($V$) is large (small) on the broken line 
in Fig. \ref{fig_U-V-phase_lambda0}, the spin susceptibility 
is dominant as compared to the charge one. 
In the region where $U$ and $V$ are intermediate on the broken line 
in Fig. \ref{fig_U-V-phase_lambda0}, the spin and charge susceptibilities 
are comparable. 
In the region where $U$ ($V$) is small (large) on the broken line 
in Fig. \ref{fig_U-V-phase_lambda0}, the charge susceptibility 
is dominant as compared to the spin one. 
Thus, surveying from point A to B along the broken line 
in Fig. \ref{fig_U-V-phase_lambda0}, dominant fluctuation changes 
from spin one to charge one. 
\par
In the region where $U$ ($V$) is large (small), 
dominant spin susceptibility, which is repulsive for spin-singlet 
pairing channel, generates spin-singlet $d_{x^2-y^2}$-wave pairing symmetry. 
This pairing symmetry has the sign change on the Fermi surface 
through the nesting vector, where the spin susceptibility has peak in 
momentum space. 
Thus, for spin-singlet $d_{x^2-y^2}$-wave pairing symmetry, 
the spin susceptibility works attractively. 
In the region where $U$ and $V$ are intermediate, 
spin-triplet pairing channel exceeds 
spin-singlet one in the pairing interaction and spin-triplet $f$-wave 
pairing symmetry is favored. 
The reason why spin-triplet pairing channel exceeds spin-singlet one is 
as follows. 
In spin-singlet pairing channel, the spin and charge susceptibilities are 
competitive in the pairing interaction. 
In spin-triplet pairing channel, on the other hand, the spin and charge 
susceptibilities are cooperative. 
In addition, since the pairing interaction for spin-triplet channel is 
originally attractive, stable pairing symmetry is $f$-wave one, which 
has no sign change on the Fermi surface through the nesting vector, 
where the spin and charge susceptibilities have peaks in momentum space. 
In the region where $U$ ($V$) is small (large), 
spin-singlet pairing channel and spin-triplet one are comparable 
in the pairing interaction. 
Then, the charge susceptibility is attractive for both pairing channels and 
spin-singlet $d_{xy}$-wave pairing symmetry is favored. 
This pairing symmetry has no sign change on the Fermi surface through 
the nesting vector, where the charge susceptibility has peak in momentum 
space. 
The total number of nodes in the gap function for spin-singlet $d_{xy}$-wave 
pairing symmetry is less than that for spin-triplet $f$-wave one. 
Details have been discussed by Onari $et$ $al$. \cite{Onari}
\subsubsection{In the presence of the RSOC}
\begin{figure}[htbp]
\includegraphics[width=0.99\linewidth,keepaspectratio]
                  {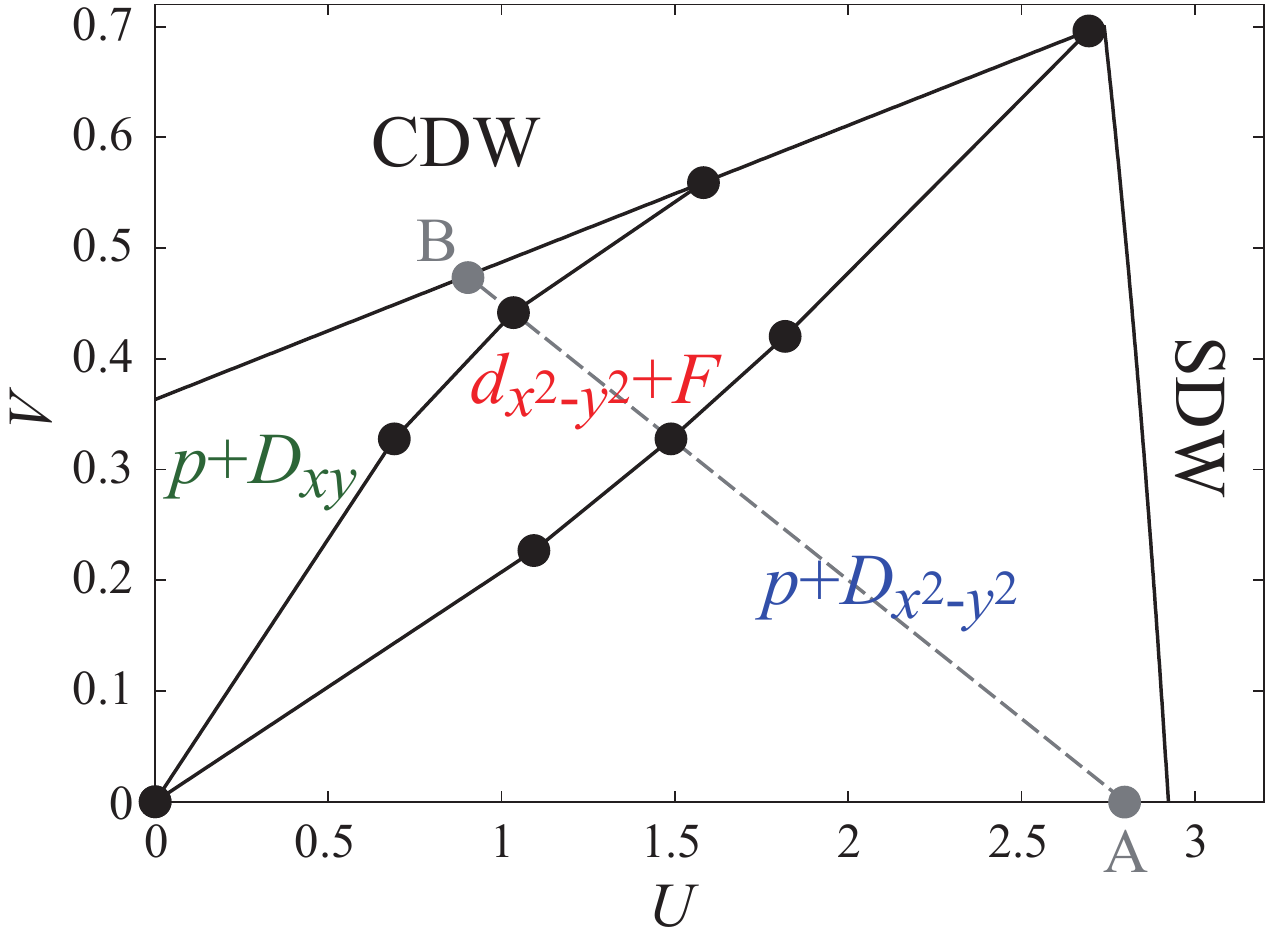}
 \caption{(Color online) $U$-$V$ phase diagram in the presence of the RSOC ($\lambda=0.3$) within the RPA. The broken line shows $V=-(U-2.8)/4$ between points A $(2.8,0)$ and B $(0.9,0.475)$.}
\label{fig_U-V-phase_lambda03}
\end{figure}

\begin{figure}[htbp]
\includegraphics[width=0.9\linewidth,keepaspectratio]
                  {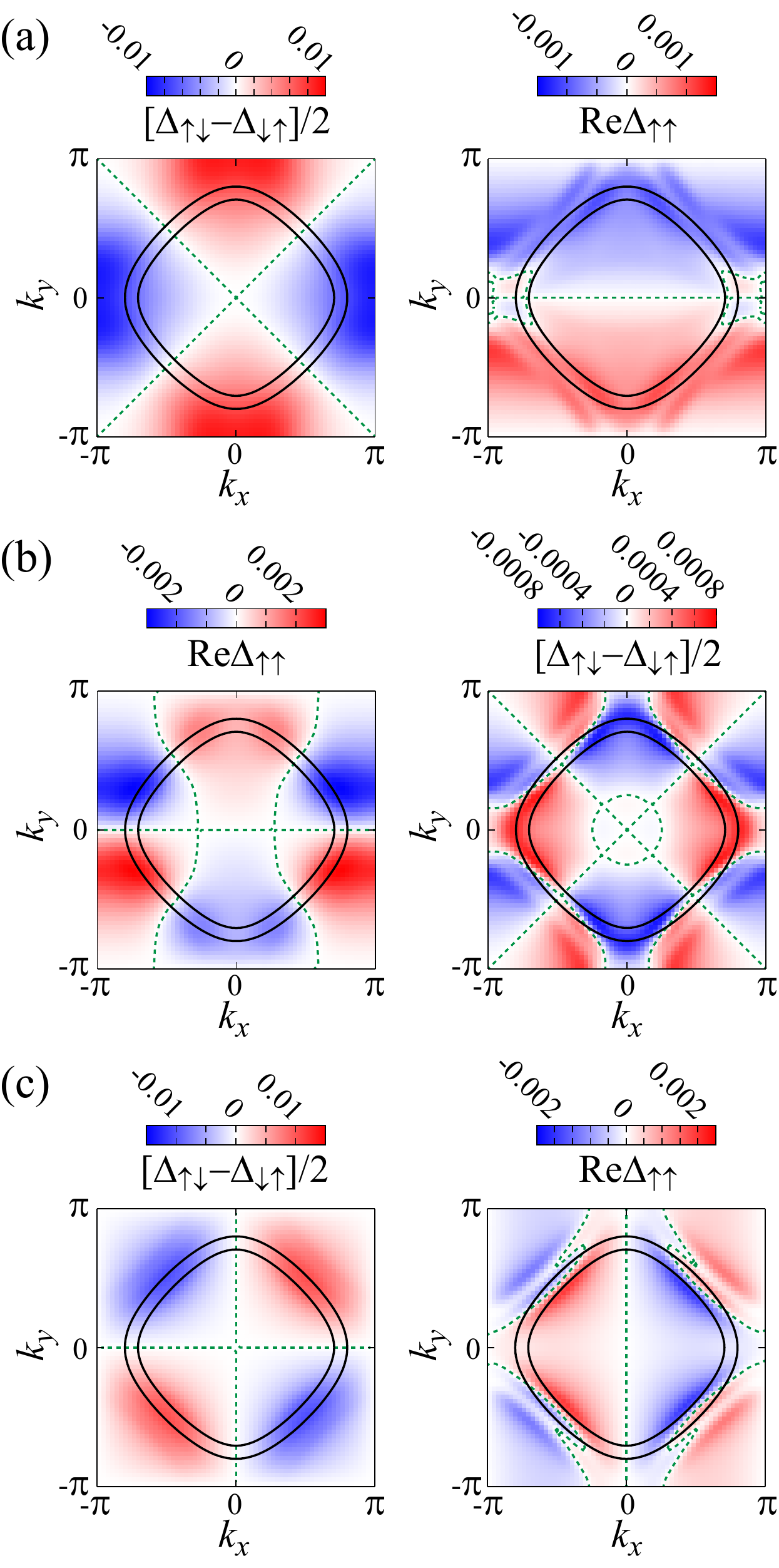}
 \caption{(Color online) Momentum dependence of the gap functions at $\omega_n=\pi T$ in the presence of the RSOC ($\lambda=0.3$) within the RPA. (a) $p+D_{x^2-y^2}$-wave pairing symmetry for $(U,V)=(2.8,0)$, (b) $d_{x^2-y^2}+F$-wave one for $(U,V)=(2.1,0.6)$, and (c) $p+D_{xy}$-wave one for $(U,V)=(0.8,0.45)$ are exhibited. Panels on the left (right) side show the predominant components (subcomponents). Black solid lines and green broken ones denote the Fermi surface and node of the gap functions, respectively. For spin-triplet components, only ${\mathrm{Re}}\Delta_{\uparrow\uparrow}(k)$ is shown.}
\label{fig_gap_lambda03}
\end{figure}

Next, we introduce the RSOC. 
It has been known that the RSOC makes the spin susceptibility anisotropic, 
$i.e.$ $\chi_{\mathrm{sp}}^{xx}(q)\neq\chi_{\mathrm{sp}}^{yy}(q)\neq\chi_{\mathrm{sp}}^{zz}(q)$ 
and $\chi_{\mathrm{sp}}^{\xi\eta}(q)\neq0$ for $\xi\neq\eta$. \cite{chi}
In the presence of the RSOC ($\lambda=0.3$), the pairing symmetry is 
shown in Fig. \ref{fig_U-V-phase_lambda03}. 
The pairing symmetry and boundary with the SDW or CDW phase are determined 
in the same manner as in the absence of the RSOC. 
As shown in Fig. \ref{fig_U-V-phase_lambda03}, 
three types of pairing symmetries can appear by tuning $U$ and $V$. 
In the region where $U$ ($V$) is large (small), 
predominantly spin-singlet $d_{x^2-y^2}$-wave pairing symmetry 
admixed with spin-triplet ($S_z=\pm1$) $p$-wave one is the most stable. 
We call this pairing symmetry $p+D_{x^2-y^2}$-wave one in the present paper. 
Momentum dependence of the gap function is shown 
in Fig. \ref{fig_gap_lambda03} (a). 
In the region where $U$ and $V$ are intermediate, 
predominantly spin-triplet ($S_z=\pm1$) $f$-wave pairing symmetry 
admixed with spin-singlet $d_{x^2-y^2}$-wave one is the most stable. 
We call this pairing symmetry $d_{x^2-y^2}+F$-wave one in the present paper. 
Momentum dependence of the gap function is shown 
in Fig. \ref{fig_gap_lambda03} (b). 
In the region where $U$ ($V$) is small (large), 
predominantly spin-singlet $d_{xy}$-wave pairing symmetry 
admixed with spin-triplet ($S_z=\pm1$) $p$-wave one is the most stable. 
We call this pairing symmetry $p+D_{xy}$-wave one in the present paper. 
Momentum dependence of the gap function is shown 
in Fig. \ref{fig_gap_lambda03} (c). 
Focusing on the predominant components, the pairing symmetries are 
the same as in the absence of the RSOC, 
$i.e.$ spin-singlet $d_{x^2-y^2}$-wave, 
spin-triplet $f$-wave, and spin-singlet $d_{xy}$-wave pairing symmetries, 
except lack of spin-triplet component with $S_z=0$. 
Phase boundaries are almost unchanged by the RSOC as shown 
in Figs. \ref{fig_U-V-phase_lambda0} and \ref{fig_U-V-phase_lambda03}. 
Namely, in the presence of the RSOC, subcomponents are admixed with 
predominant components whose pairing symmetries are determined by 
cooperative/competitive spin and charge fluctuations without 
the RSOC as discussed above. 
It is discussed later in \S\ref{sec_analytical} how the pairing symmetries 
of the subcomponents are determined. 
\par
\begin{figure}[htbp]
\includegraphics[width=0.99\linewidth,keepaspectratio]
                  {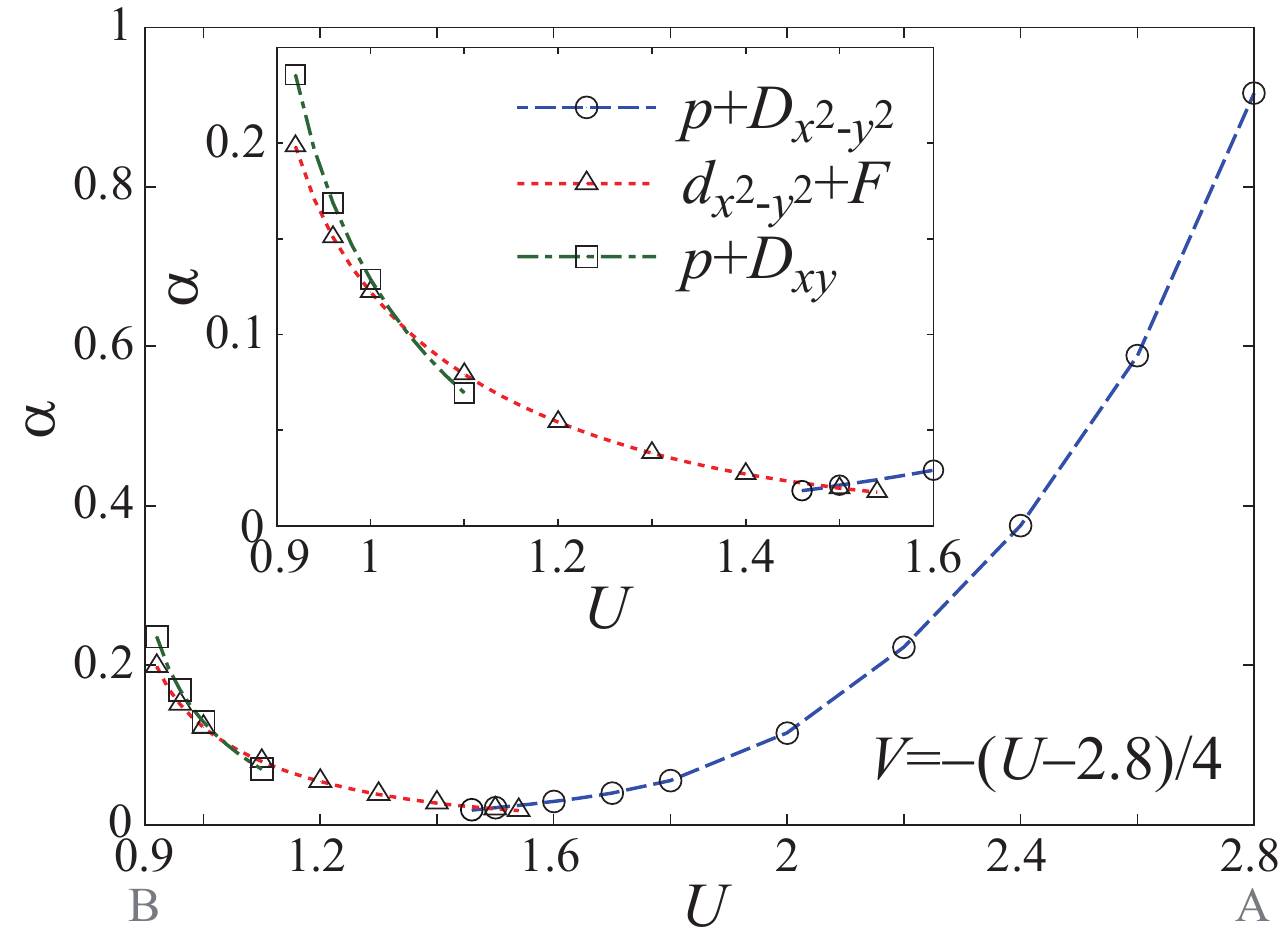}
 \caption{(Color online) The eigenvalues $\alpha$ in the linearized ${\acute{\mathrm{E}}}$liashberg's equation (\ref{eq_eliash1}) and (\ref{eq_eliash2}) on the broken line $V=-(U-2.8)/4$ in Fig. \ref{fig_U-V-phase_lambda03} ($U$-$V$ space) with $\lambda=0.3$. The inset is a enlarged view of the region where $U$ ($V$) is small (large). The result is numerically obtained within the RPA. Points A and B correspond to those in Fig. \ref{fig_U-V-phase_lambda03}.}
\label{fig_U_alpha_lambda03}
\end{figure}

The eigenvalues $\alpha$ in the linearized 
${\acute{\mathrm{E}}}$liashberg's equation 
(\ref{eq_eliash1}) and (\ref{eq_eliash2}) change with $U$ and $V$ 
in the presence of the RSOC ($\lambda=0.3$) as shown 
in Fig. \ref{fig_U_alpha_lambda03}, 
where the data on the broken line $V=-(U-2.8)/4$ 
in Fig. \ref{fig_U-V-phase_lambda03} ($U$-$V$ space) are exhibited. 
In the vicinity of the boundary with the SDW or CDW phase, 
the eigenvalue $\alpha$ is large. 
Away from the boundary, the eigenvalue $\alpha$ 
is small. 
However, even away from the boundary, the eigenvalue $\alpha$ is expected 
to increase with decreasing temperature. 
\par
\begin{figure}[htbp]
\includegraphics[width=0.99\linewidth,keepaspectratio]
                  {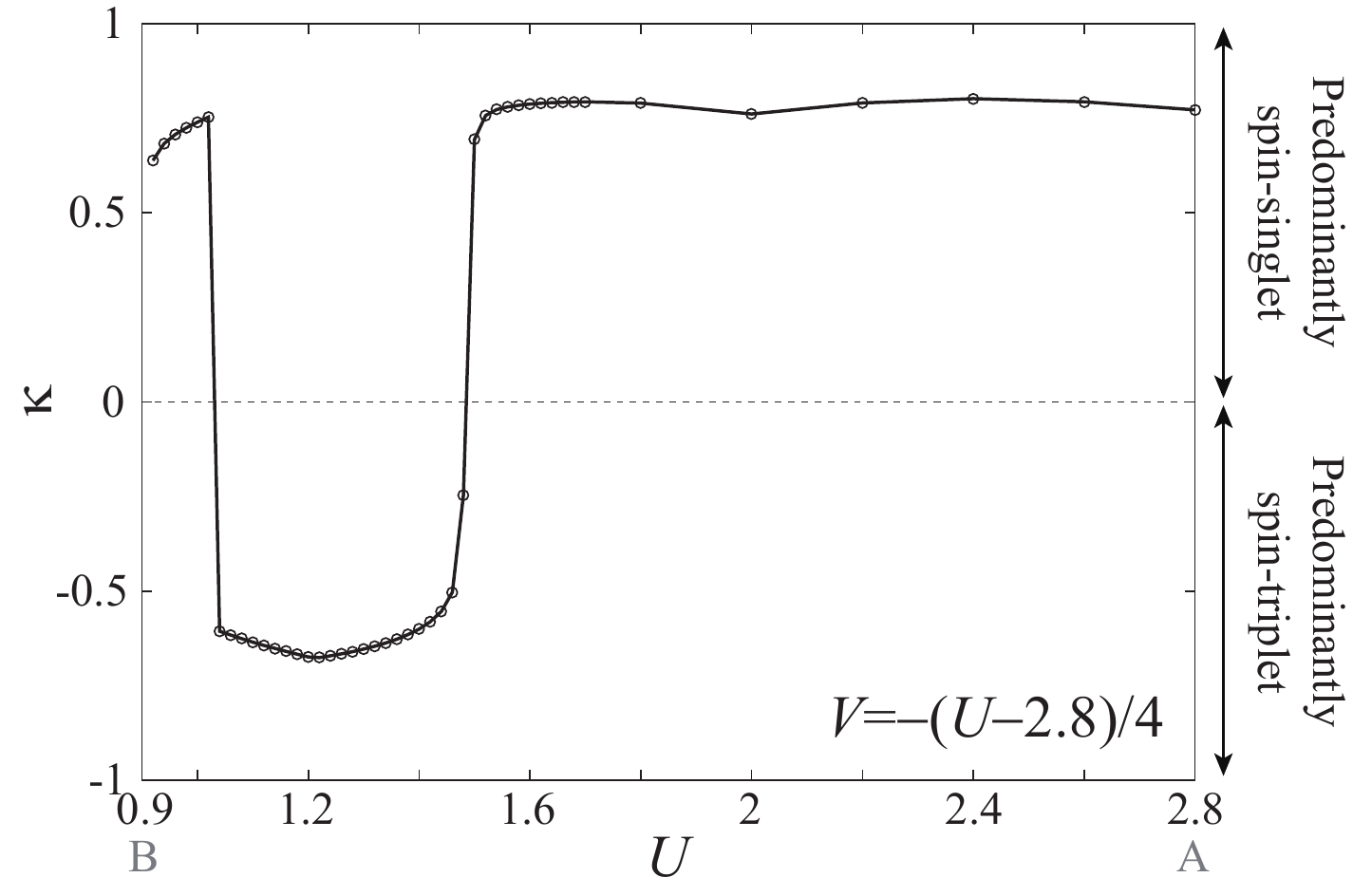}
 \caption{Ratio $\kappa$ between spin-singlet and -triplet ($S_{z}=\pm1$) components of the gap function on the broken line $V=-(U-2.8)/4$ in Fig. \ref{fig_U-V-phase_lambda03} ($U$-$V$ space) with $\lambda=0.3$. The result is numerically obtained within the RPA. Points A and B correspond to those in Fig. \ref{fig_U-V-phase_lambda03}.}
\label{fig_U_ratio_lambda03}
\end{figure}

We also look at ratio between spin-singlet and -triplet ($S_{z}=\pm1$) 
components of the gap function on the broken line $V=-(U-2.8)/4$ 
in Fig. \ref{fig_U-V-phase_lambda03} ($U$-$V$ space). 
Fig. \ref{fig_U_ratio_lambda03} shows the ratio, which is defined by 
$\kappa\equiv\left[\Delta_{\mathrm{s}}-\Delta_{\mathrm{t}}\right]/\left[\Delta_{\mathrm{s}}+\Delta_{\mathrm{t}}\right]$. 
Here, $\Delta_{\mathrm{s}}$ and $\Delta_{\mathrm{t}}$ denote 
the maximum absolute values of spin-singlet and -triplet ($S_{z}=\pm1$) 
components, respectively. 
$\kappa=1$ corresponds to purely spin-singlet pairing state while 
$\kappa=-1$ corresponds to purely spin-triplet one. 
As shown in Fig. \ref{fig_U_ratio_lambda03}, the ratio $\kappa$ slightly 
depends on $U$ and $V$ in each region, $i.e.$ 
$p+D_{x^2-y^2}$-wave pairing region ($0.92\lesssim U\lesssim1.03$), 
$d_{x^2-y^2}+F$-wave pairing one ($1.03\lesssim U\lesssim1.48$), and 
$p+D_{xy}$-wave pairing one ($1.48\lesssim U\leq2.8$). 
Namely, change of pairing interaction slightly affects the ratio $\kappa$ 
in each region. 
The jump of $\kappa$ on the phase boundaries indicates that 
phase transitions between different pairing symmetries are not 
crossover even in the presence of the RSOC. 
\par
\begin{figure}[htbp]
\includegraphics[width=0.99\linewidth,keepaspectratio]
                  {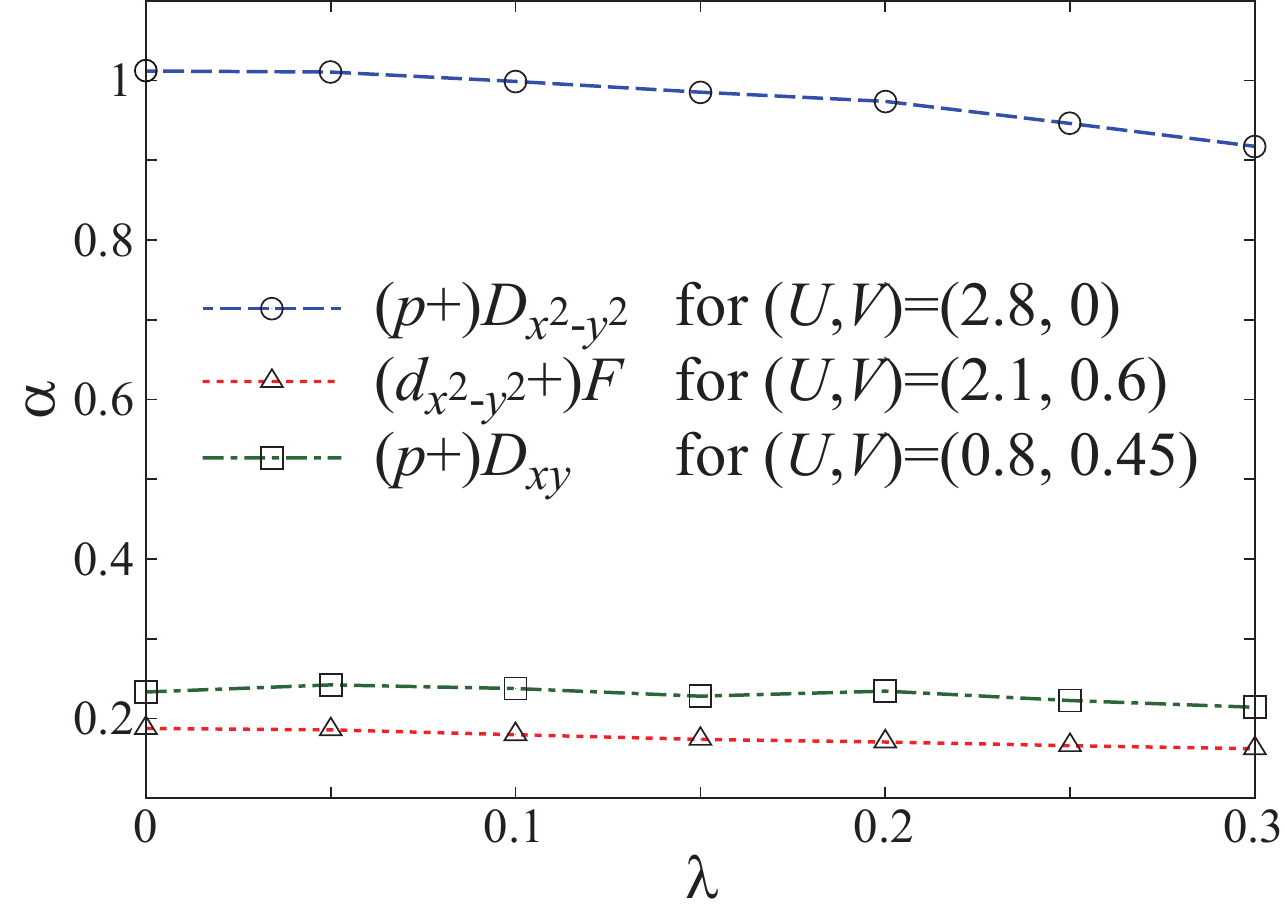}
 \caption{(Color online) The eigenvalue $\alpha$ in the linearized ${\acute{\mathrm{E}}}$liashberg's equation (\ref{eq_eliash1}) and (\ref{eq_eliash2}) as a function of $\lambda$ for $(p+)D_{x^2-y^2}$-wave pairing symmetry at $(U,V)=(2.8,0)$, $(d_{x^2-y^2}+)F$-wave one at $(U,V)=(2.1,0.6)$, and $(p+)D_{xy}$-wave one at $(U,V)=(0.8,0.45)$. The result is numerically obtained within the RPA.}
\label{fig_r_alpha}
\end{figure}
\begin{figure}[htbp]
\includegraphics[width=0.99\linewidth,keepaspectratio]
                  {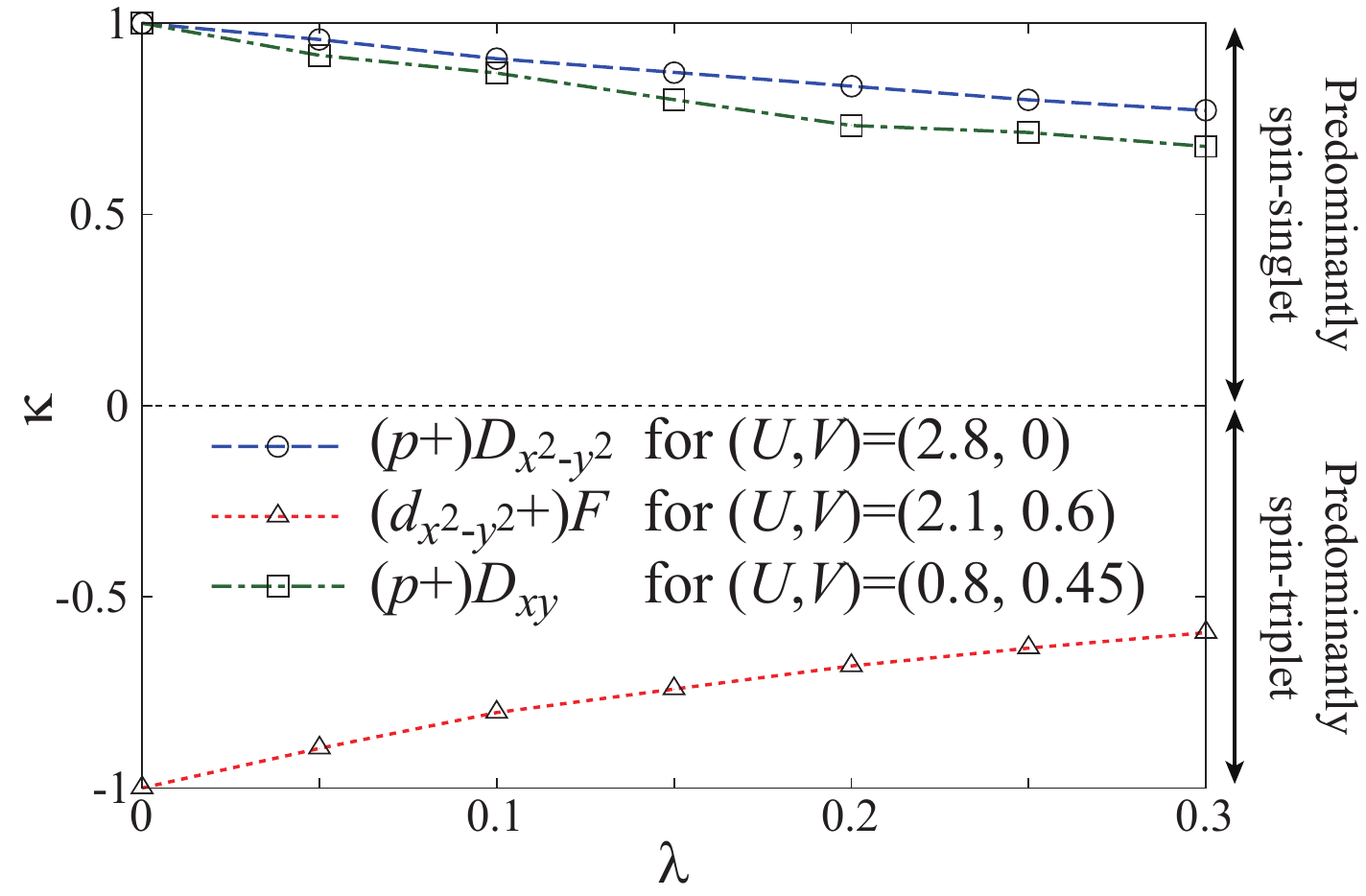}
 \caption{(Color online) Ratio $\kappa$ between spin-singlet and -triplet ($S_{z}=\pm1$) components of the gap function as a function of $\lambda$ for $(p+)D_{x^2-y^2}$-wave pairing symmetry at $(U,V)=(2.8,0)$, $(d_{x^2-y^2}+)F$-wave one at $(U,V)=(2.1,0.6)$, and $(p+)D_{xy}$-wave one at $(U,V)=(0.8,0.45)$. The result is numerically obtained within the RPA.}
\label{fig_r_ratio}
\end{figure}

As functions of $\lambda$, the eigenvalue $\alpha$ in the linearized 
${\acute{\mathrm{E}}}$liashberg's equation 
(\ref{eq_eliash1}) and (\ref{eq_eliash2}) and the ratio $\kappa$ between 
spin-singlet and -triplet ($S_{z}=\pm1$) components of the gap function 
are shown in Figs. \ref{fig_r_alpha} and \ref{fig_r_ratio}, respectively. 
In both figures, data for 
$(p+)D_{x^2-y^2}$-wave pairing symmetry at $(U,V)=(2.8,0)$, 
$(d_{x^2-y^2}+)F$-wave one at $(U,V)=(2.1,0.6)$, and 
$(p+)D_{xy}$-wave one at $(U,V)=(0.8,0.45)$ are exhibited. 
As shown in Fig. \ref{fig_r_alpha}, the eigenvalue $\alpha$ slightly 
decreases with increasing $\lambda$ because the split Fermi surface 
makes the nesting condition worse. 
As shown in Fig. \ref{fig_r_ratio}, the admixture of spin-singlet and -triplet 
($S_{z}=\pm1$) components is enhanced by $\lambda$ linearly as discussed 
later in \S\ref{sec_analytical}. 
\par
In the above discussion, we mention the pairing symmetry 
in the presence of the RSOC with 
focusing on only the predominant component and subcomponent whose amplitude 
is secondarily large in the gap function. 
Those always correspond to spin-singlet and -triplet ($S_z=\pm1$) pairings. 
However, we also obtain a spin-triplet ($S_z=0$) component 
whose amplitude is small as compared to those of the above components. 
The spin-triplet ($S_z=0$) 
component is always an odd function with respect to the Matsubara frequency, 
so-called odd-frequency pairing state. 
\cite{ext_Hub11,Berezinskii,odd,odd2,odd3,odd4,odd5,odd6,odd7,odd8,odd9,odd10,odd11,odd12,odd13,odd14,odd15,odd16,odd17,odd18,odd19,odd20} 
Since Berezinskii proposed the odd-frequency pairing state in 1974, 
\cite{Berezinskii} 
it has been an important issue in superconductivity/superfluidity. 
It is expected that the odd-frequency pairing state is discovered 
in non-centrosymmetric superconductors. 
\subsection{Analytical discussion}
\label{sec_analytical}
In this subsection, we discuss the pairing symmetry in the presence of 
the RSOC analytically. 
In the previous subsection, it has been clarified that 
pairing symmetry of the predominant component of the gap function 
in the presence of the RSOC is the same as in the absence of the RSOC. 
Now, we focus on pairing symmetry of the admixed subcomponent 
in the presence of the RSOC. 
We discuss below how the pairing symmetry of the subcomponent is 
determined for each predominant pairing symmetry, $i.e.$ 
spin-singlet $d_{x^2-y^2}$-wave one, spin-triplet ($S_z=\pm1$) $f$-wave one, 
and spin-singlet $d_{xy}$-wave one. 
\par
According to the linearized ${\acute{\mathrm{E}}}$liashberg's equation 
(\ref{eq_eliash2}), if a spin-singlet component is predominant, 
an admixed spin-triplet ($S_z=1$) subcomponent of the anomalous Green's 
function is given by 
\begin{align}
F_{\uparrow\uparrow}^{\mathrm{sub}}(k)=
\frac{\lambda\varepsilon_{\boldsymbol{k}}\left[-g_x(\boldsymbol{k})+{\mathrm{i}}g_y(\boldsymbol{k})\right]}{(\omega_n^2+\varepsilon_{\boldsymbol{k}}^2)^2}
\left[\Delta_{\uparrow\downarrow}^{\mathrm{dom}}(k)-\Delta_{\downarrow\uparrow}^{\mathrm{dom}}(k)\right],
\label{eq_F_t}
\end{align}
with expanding the Green's function up to the first order with respect to 
$\lambda$. 
Here, $F_{ss'}^{\mathrm{sub}}(k)$ and $\Delta_{ss'}^{\mathrm{dom}}(k)$ 
denote the admixed subcomponent of the anomalous Green's function and 
the predominant component of the gap function, respectively. 
Similarly, if a spin-triplet ($S_z=\pm1$) component is predominant, 
an admixed spin-singlet subcomponent of the anomalous Green's function is 
given by 
\begin{align}
&F_{\uparrow\downarrow}^{\mathrm{sub}}(k)-F_{\downarrow\uparrow}^{\mathrm{sub}}(k) \notag \\
&=\frac{4\lambda\varepsilon_{\boldsymbol{k}}}{(\omega_n^2+\varepsilon_{\boldsymbol{k}}^2)^2}
\left[-g_x({\boldsymbol{k}}){\mathrm{Re}}\Delta_{\uparrow\uparrow}^{\mathrm{dom}}(k)+g_y({\boldsymbol{k}}){\mathrm{Im}}\Delta_{\uparrow\uparrow}^{\mathrm{dom}}(k)\right].
\label{eq_F_s}
\end{align}
Eqs. (\ref{eq_F_t}) and (\ref{eq_F_s}) indicate that amplitude of an admixed 
subcomponent increases linearly with $\lambda$. 
This is consistent with the numerical result within the RPA shown 
in Fig. \ref{fig_r_ratio}. 
\par
From the anomalous Green's function, we can derive the gap function by 
using the linearized ${\acute{\mathrm{E}}}$liashberg's equation 
(\ref{eq_eliash1}). 
For simplicity, we approximately apply $P^c({\boldsymbol{k}})=0$ for 
$c\neq1$ hereafter because the pairing symmetry is mainly determined by 
the term with $P^1({\boldsymbol{k'}})P^1(-{\boldsymbol{k}})=1$ in 
the linearized ${\acute{\mathrm{E}}}$liashberg's equation (\ref{eq_eliash1}) 
in the actual calculation within the RPA. 
Then, using the convolution theorem, 
the linearized ${\acute{\mathrm{E}}}$liashberg's equation (\ref{eq_eliash1}) 
is rewritten as 
\begin{align}
\Delta_{ss'}^{\mathrm{sub}}(k)=&-\frac{T}{N}{\cal{F}}^{-1}\{\bar{\Gamma}_{ss'ss'}^{11}(r)\bar{F}_{ss'}^{\mathrm{sub}}(r)\}, 
\label{eq_convolution}
\\
\bar{\Gamma}_{ss'ss'}^{11}(r)=&{\cal{F}}\{\Gamma_{ss'ss'}^{11}(q)\}, \\
\bar{F}_{ss'}^{\mathrm{sub}}(r)=&{\cal{F}}\{F_{ss'}^{\mathrm{sub}}(k)\}, 
\label{eq_fourier}
\end{align}
where ${\cal{F}}^{(-1)}$ denotes (inverse) Fourier transformation. 
In the present subsection, we focus on the pairing symmetry, 
$i.e.$ the nodal structure of the gap function. 
From this viewpoint, it is available to neglect factors which generate 
no node in the anomalous Green's function, $i.e.$ 
$\lambda/(\omega_n^2+\varepsilon_{\boldsymbol{k}}^2)^2$ in Eq. (\ref{eq_F_t}) 
and $4\lambda/(\omega_n^2+\varepsilon_{\boldsymbol{k}}^2)^2$ 
in Eq. (\ref{eq_F_s}). 
Thus, we can rewrite the anomalous 
Green's functions as follows, 
\begin{align}
&\tilde{F}_{\uparrow\uparrow}^{\mathrm{sub}}(k)=
(-2t\cos k_x-2t\cos k_y-\mu)(\sin k_y+{\mathrm{i}}\sin k_x) \notag \\
&\hspace{16mm}\times\left[\Delta_{\uparrow\downarrow}^{\mathrm{dom}}(k)-\Delta_{\downarrow\uparrow}^{\mathrm{dom}}(k)\right],
\label{eq_F_t_2} \\
&\tilde{F}_{\uparrow\downarrow}^{\mathrm{sub}}(k)-\tilde{F}_{\downarrow\uparrow}^{\mathrm{sub}}(k) \notag \\
&\hspace{12mm}=(-2t\cos k_x-2t\cos k_y-\mu) \notag \\
&\hspace{16mm}\times\left[\sin k_y{\mathrm{Re}}\Delta_{\uparrow\uparrow}^{\mathrm{dom}}(k)+\sin k_x{\mathrm{Im}}\Delta_{\uparrow\uparrow}^{\mathrm{dom}}(k)\right],
\label{eq_F_s_2}
\end{align}
which are effective in deriving the nodal structure of the gap function, 
by actually substituting $\varepsilon_{\boldsymbol{k}}$ and 
${\boldsymbol{g}}({\boldsymbol{k}})$ 
in Eqs. (\ref{eq_F_t}) and (\ref{eq_F_s}). 
Here, we neglect $t'$ for simplicity. 
By using $\tilde{F}_{\uparrow\uparrow}^{\mathrm{sub}}(k)$ and 
$\tilde{F}_{\uparrow\downarrow}^{\mathrm{sub}}(k)-\tilde{F}_{\downarrow\uparrow}^{\mathrm{sub}}(k)$ 
instead of $F_{\uparrow\uparrow}^{\mathrm{sub}}(k)$ and 
$F_{\uparrow\downarrow}^{\mathrm{sub}}(k)-F_{\downarrow\uparrow}^{\mathrm{sub}}(k)$, 
respectively, in Eq. (\ref{eq_fourier}), we can derive the pairing 
symmetry of the admixed subcomponent of the gap function. 
\par
\begin{figure}[htbp]
\includegraphics[width=0.8\linewidth,keepaspectratio]
                  {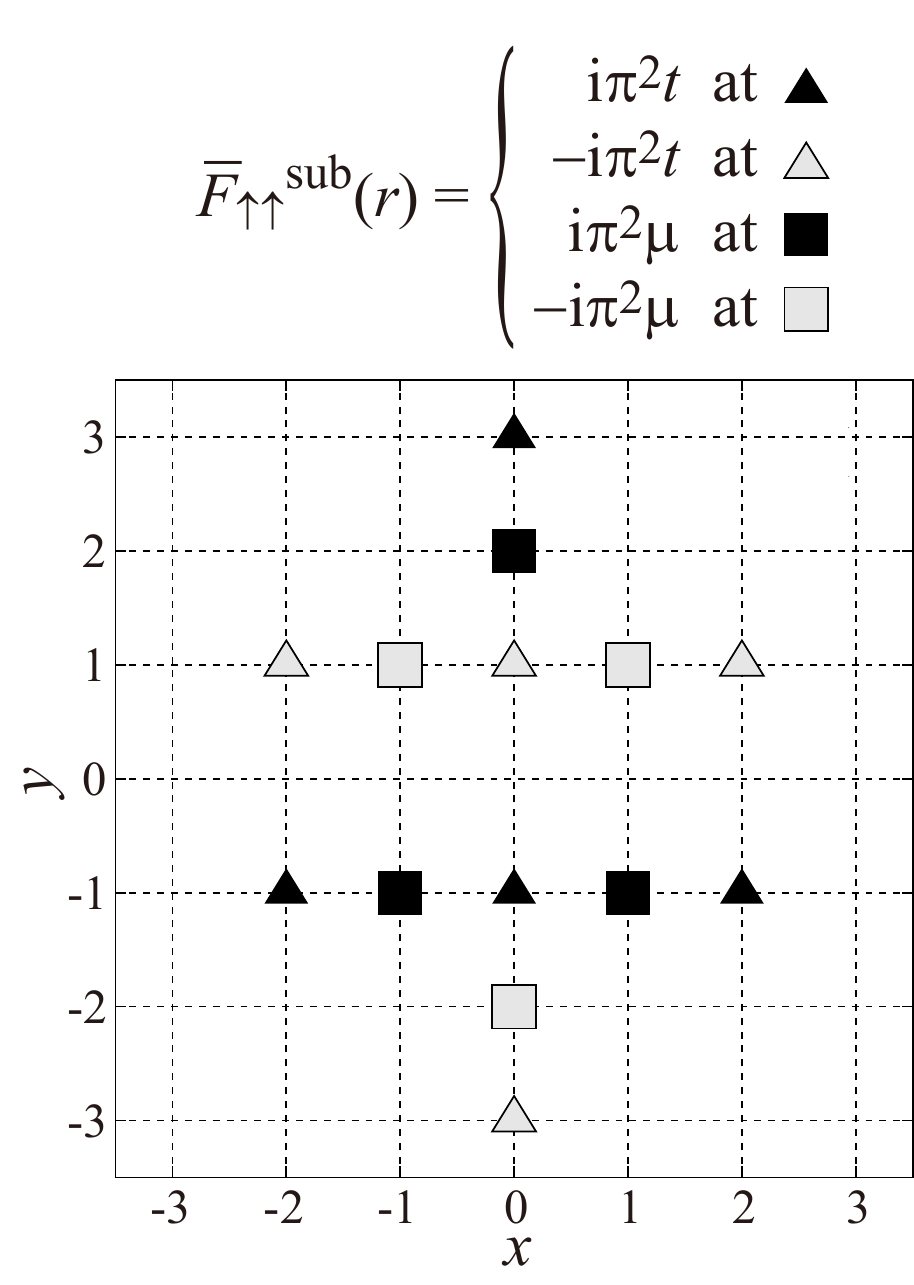}
 \caption{$\bar{F}_{\uparrow\uparrow}^{\mathrm{sub}}(r)$, which indicates the nodal structure of the admixed spin-triplet ($S_z=1$) subcomponent of the anomalous Green's function (\ref{eq_F_t}), for given predominant spin-singlet $d_{x^2-y^2}$-wave component approximated by $\Delta_{\uparrow\downarrow}^{\mathrm{dom}}(k)-\Delta_{\downarrow\uparrow}^{\mathrm{dom}}(k)=\cos k_x-\cos k_y$. Note that, strictly speaking, $\bar{F}_{\uparrow\uparrow}^{\mathrm{sub}}(r)$ displayed here corresponds to ${\cal{F}}\{{\mathrm{Re}}\tilde{F}_{\uparrow\uparrow}^{\mathrm{sub}}(k)\}$.}
\label{fig_F1}
\end{figure}

With use of the above analytical discussion, we first discuss 
$p+D_{x^2-y^2}$-wave pairing symmetry in the region where $U$ ($V$) is 
large (small). 
In this region, the dominant spin susceptibility mediates predominant 
spin-singlet $d_{x^2-y^2}$-wave pairing. 
We approximate the gap function by 
$\Delta_{\uparrow\downarrow}^{\mathrm{dom}}(k)-\Delta_{\downarrow\uparrow}^{\mathrm{dom}}(k)=\cos k_x-\cos k_y$. 
For the given predominant component, we can calculate 
$\tilde{F}_{\uparrow\uparrow}^{\mathrm{sub}}(k)$ 
by using Eq. (\ref{eq_F_t_2}). 
For simplicity, we consider only the real part in 
$\tilde{F}_{\uparrow\uparrow}^{\mathrm{sub}}(k)$ below. 
Fig. \ref{fig_F1} shows $\bar{F}_{\uparrow\uparrow}^{\mathrm{sub}}(r)$ 
obtained by Fourier transform (\ref{eq_fourier}). 
Note that this $\bar{F}_{\uparrow\uparrow}^{\mathrm{sub}}(r)$ 
corresponds to 
${\cal{F}}\{{\mathrm{Re}}\tilde{F}_{\uparrow\uparrow}^{\mathrm{sub}}(k)\}$. 
There are two kinds of spots in real space. 
One is the spot whose amplitude increases with $\left| t\right|$, 
which actually depends on materials while it is chosen as a unit of energy 
in \S\ref{sec_RPA}. 
The other is the spot whose amplitude increases with $\left|\mu\right|$. 
Namely, the pairing symmetry of the admixed subcomponent of the gap function 
is determined by not only that of the predominant component 
but also the dispersion relation. 
On the other hand, the pairing interaction 
$\bar{\Gamma}_{\uparrow\uparrow\uparrow\uparrow}^{11}(r)$ 
for the admixed spin-triplet ($S_z=1$) subcomponent 
has large amplitude at $(x,y)=(0,0),(\pm1,0),(0,\pm1)$ in this $(U,V)$ region 
within the RPA. 
The value at $(x,y)=(0,0)$ is negative, $i.e.$ attractive, 
while the values at $(x,y)=(\pm1,0),(0,\pm1)$ are positive, $i.e.$ repulsive. 
Then, the resulting product 
$\bar{\Gamma}_{\uparrow\uparrow\uparrow\uparrow}^{11}(r)\bar{F}_{\uparrow\uparrow}^{\mathrm{sub}}(r)$ 
in Eq. (\ref{eq_convolution}) has values at $(x,y)=(0,\pm1)$. 
Sign of the value at $(x,y)=(0,1)$ is opposite to that at $(x,y)=(0,-1)$. 
This amplitude increases with $\left| t\right|$ at $(x,y)=(0,\pm1)$ 
as shown in Fig. \ref{fig_F1}. 
Thus, the admixed spin-triplet ($S_z=1$) subcomponent becomes $p$-wave with 
$\Delta_{\uparrow\uparrow}^{\mathrm{sub}}(k)\propto\sin k_y$. 
Note that the above $\Delta_{\uparrow\uparrow}^{\mathrm{sub}}(k)$, which 
is derived from the real part of 
$\tilde{F}_{\uparrow\uparrow}^{\mathrm{sub}}(k)$, is real. 
The imaginary part of $\tilde{F}_{\uparrow\uparrow}^{\mathrm{sub}}(k)$ 
gives the imaginary part of $\Delta_{\uparrow\uparrow}^{\mathrm{sub}}(k)$, 
which is also $p$-wave. 
By the similar procedure, it is derived that 
$\Delta_{\downarrow\downarrow}^{\mathrm{sub}}(k)$ is also $p$-wave. 
Thus, the admixed spin-triplet ($S_z=\pm1$) subcomponent is found 
to be $p$-wave and the amplitude increases with $\left| t\right|$. 
\par
\begin{figure}[htbp]
\includegraphics[width=0.86\linewidth,keepaspectratio]
                  {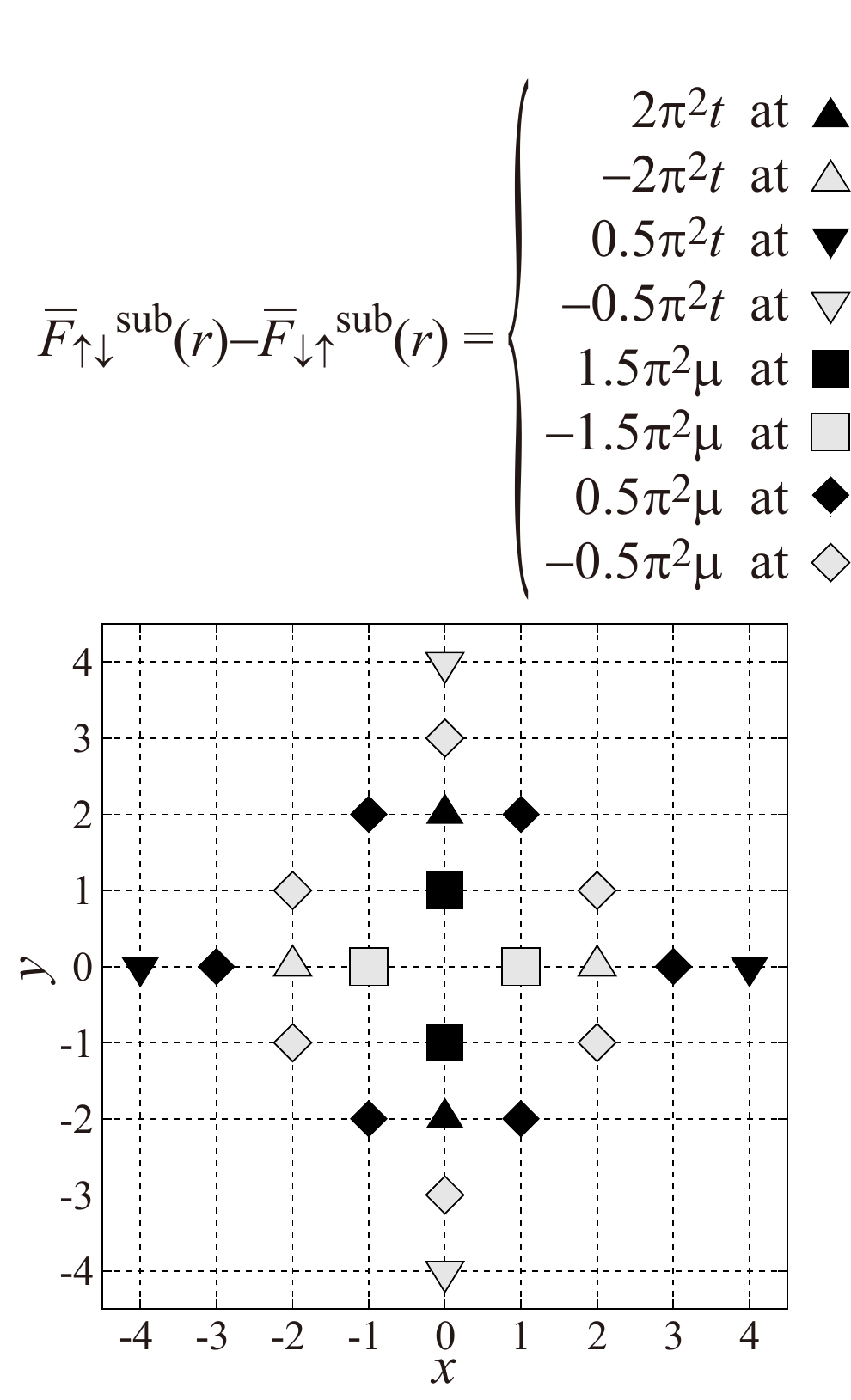}
 \caption{$\bar{F}_{\uparrow\downarrow}^{\mathrm{sub}}(r)-\bar{F}_{\downarrow\uparrow}^{\mathrm{sub}}(r)$, which indicates the nodal structure of the admixed spin-singlet subcomponent of the anomalous Green's function (\ref{eq_F_s}), for given predominant spin-triplet $f$-wave component approximated by $\Delta_{\uparrow\uparrow}^{\mathrm{dom}}(k)=(\cos k_x-\cos k_y)\sin k_y+{\mathrm{i}}(\cos k_x-\cos k_y)\sin k_x$.}
\label{fig_F2}
\end{figure}

Next, we discuss $d_{x^2-y^2}+F$-wave pairing symmetry in the region 
where $U$ and $V$ are intermediate. 
In this region, the comparable spin and charge susceptibilities mediate 
predominant spin-triplet $f$-wave pairing. 
We approximate the gap function by 
$\Delta_{\uparrow\uparrow}^{\mathrm{dom}}(k)=(\cos k_x-\cos k_y)\sin k_y+{\mathrm{i}}(\cos k_x-\cos k_y)\sin k_x$. 
For the given predominant component, we can calculate 
$\tilde{F}_{\uparrow\downarrow}^{\mathrm{sub}}(k)-\tilde{F}_{\downarrow\uparrow}^{\mathrm{sub}}(k)$ 
by using Eq. (\ref{eq_F_s_2}). 
Fig. \ref{fig_F2} shows 
$\bar{F}_{\uparrow\downarrow}^{\mathrm{sub}}(r)-\bar{F}_{\downarrow\uparrow}^{\mathrm{sub}}(r)$ 
obtained by Fourier transform (\ref{eq_fourier}). 
On the other hand, the pairing interaction 
$\bar{\Gamma}_{\uparrow\downarrow\uparrow\downarrow}^{11}(r)=\bar{\Gamma}_{\downarrow\uparrow\downarrow\uparrow}^{11}(r)$ 
for the admixed spin-singlet subcomponent 
has large amplitude at $(x,y)=(0,0),(\pm1,0),(0,\pm1)$ in this $(U,V)$ region 
within the RPA. 
All the values are positive, $i.e.$ repulsive. 
Then, the resulting product 
$\bar{\Gamma}_{\uparrow\downarrow\uparrow\downarrow}^{11}(r)\left[\bar{F}_{\uparrow\downarrow}^{\mathrm{sub}}(r)-\bar{F}_{\downarrow\uparrow}^{\mathrm{sub}}(r)\right]$ 
in Eq. (\ref{eq_convolution}) has values at $(x,y)=(\pm1,0)$ and 
$(x,y)=(0,\pm1)$. 
Signs of the values at $(x,y)=(\pm1,0)$ are opposite to those at 
$(x,y)=(0,\pm1)$. 
This amplitude increases with $\left|\mu\right|$ at $(x,y)=(\pm1,0),(0,\pm1)$ 
as shown in Fig. \ref{fig_F2}. 
Thus, the admixed spin-singlet subcomponent becomes $d_{x^2-y2}$-wave with 
$\Delta_{\uparrow\downarrow}^{\mathrm{sub}}(k)-\Delta_{\downarrow\uparrow}^{\mathrm{sub}}(k)\propto\cos k_x-\cos k_y$. 
The amplitude is found to increase with $\left|\mu\right|$. 
\par
\begin{figure}[htbp]
\includegraphics[width=0.8\linewidth,keepaspectratio]
                  {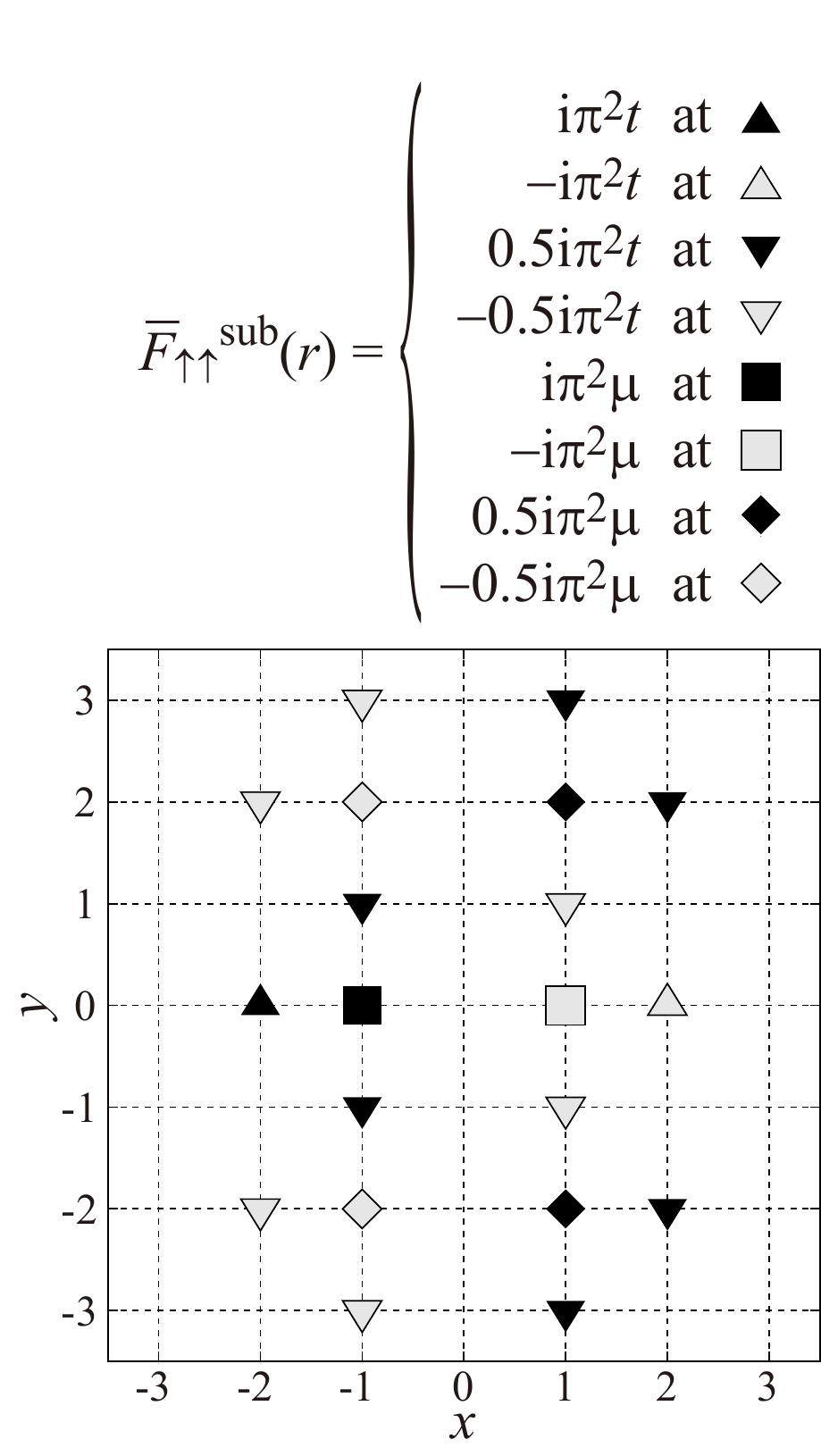}
 \caption{$\bar{F}_{\uparrow\uparrow}^{\mathrm{sub}}(r)$, which indicates the nodal structure of the admixed spin-triplet ($S_z=1$) subcomponent of the anomalous Green's function (\ref{eq_F_t}), for given predominant spin-singlet $d_{xy}$-wave component approximated by $\Delta_{\uparrow\downarrow}^{\mathrm{dom}}(k)-\Delta_{\downarrow\uparrow}^{\mathrm{dom}}(k)=\sin k_x\sin k_y$. Note that, strictly speaking, $\bar{F}_{\uparrow\uparrow}^{\mathrm{sub}}(r)$ displayed here corresponds to ${\cal{F}}\{{\mathrm{Re}}\tilde{F}_{\uparrow\uparrow}^{\mathrm{sub}}(k)\}$.}
\label{fig_F3}
\end{figure}

Finally, we discuss $p+D_{xy}$-wave pairing symmetry in the region 
where $U$ ($V$) is small (large). 
In this region, the dominant charge susceptibility mediates predominant 
spin-singlet $d_{xy}$-wave pairing. 
We approximate the gap function by 
$\Delta_{\uparrow\downarrow}^{\mathrm{dom}}(k)-\Delta_{\downarrow\uparrow}^{\mathrm{dom}}(k)=\sin k_x\sin k_y$. 
For the given predominant component, we can calculate 
$\tilde{F}_{\uparrow\uparrow}^{\mathrm{sub}}(k)$ 
by using Eq. (\ref{eq_F_t_2}). 
For simplicity, we consider only the real part in 
$\tilde{F}_{\uparrow\uparrow}^{\mathrm{sub}}(k)$ below. 
Fig. \ref{fig_F3} shows $\bar{F}_{\uparrow\uparrow}^{\mathrm{sub}}(r)$ 
obtained by Fourier transform (\ref{eq_fourier}). 
Note that this $\bar{F}_{\uparrow\uparrow}^{\mathrm{sub}}(r)$ 
corresponds to 
${\cal{F}}\{{\mathrm{Re}}\tilde{F}_{\uparrow\uparrow}^{\mathrm{sub}}(k)\}$. 
On the other hand, the pairing interaction 
$\bar{\Gamma}_{\uparrow\uparrow\uparrow\uparrow}^{11}(r)$ 
for the admixed spin-triplet ($S_z=1$) subcomponent 
has large amplitude at $(x,y)=(0,0),(\pm1,0),(0,\pm1)$ in this $(U,V)$ region 
within the RPA. 
The value at $(x,y)=(0,0)$ is negative, $i.e.$ attractive, 
while the values at $(x,y)=(\pm1,0),(0,\pm1)$ are positive, $i.e.$ repulsive. 
Then, the resulting product 
$\bar{\Gamma}_{\uparrow\uparrow\uparrow\uparrow}^{11}(r)\bar{F}_{\uparrow\uparrow}^{\mathrm{sub}}(r)$ 
in Eq. (\ref{eq_convolution}) has values at $(x,y)=(\pm1,0)$. 
Sign of the value at $(x,y)=(1,0)$ is opposite to that at $(x,y)=(-1,0)$. 
This amplitude increases with $\left|\mu\right|$ at $(x,y)=(\pm1,0)$ 
as shown in Fig. \ref{fig_F3}. 
Thus, the admixed spin-triplet ($S_z=1$) subcomponent becomes $p$-wave with 
$\Delta_{\uparrow\uparrow}^{\mathrm{sub}}(k)\propto\sin k_x$. 
Note that the above $\Delta_{\uparrow\uparrow}^{\mathrm{sub}}(k)$, which 
is derived from the real part of 
$\tilde{F}_{\uparrow\uparrow}^{\mathrm{sub}}(k)$, is real. 
The imaginary part of $\tilde{F}_{\uparrow\uparrow}^{\mathrm{sub}}(k)$ 
gives the imaginary part of $\Delta_{\uparrow\uparrow}^{\mathrm{sub}}(k)$, 
which is also $p$-wave. 
By the similar procedure, it is derived that 
$\Delta_{\downarrow\downarrow}^{\mathrm{sub}}(k)$ is also $p$-wave. 
Thus, the admixed spin-triplet ($S_z=\pm1$) subcomponent is found 
to be $p$-wave and the amplitude increases with $\left|\mu\right|$. 
\par
In the above analytical discussion, we derive the pairing symmetry 
of the admixed subcomponent of the gap function 
with use of the simplified pairing interaction, 
where some dominant modes decomposed in real space are chosen. 
The derived pairing symmetry is in agreement with that numerically 
calculated within the RPA in terms of the symmetrical class such as 
$p$- or $d$-wave symmetry. 
However, there is difference in the nodal structure in detail. 
The difference comes from the simplification of the pairing interaction. 
Considering the higher order harmonic components in the pairing interaction, 
of course, we can reproduce in detail the nodal structure 
in the admixed subcomponent of the gap function obtained by the numerical 
calculation within the RPA. 
\par
There exists previous analytical discussion on the admixture of 
the pairing symmetry. \cite{magnetoelectric4} 
According to the previous study, spin-triplet components are 
related to spin-singlet one in the gap function by 
\begin{align}
{\boldsymbol{d}}(k)\propto\frac{{\boldsymbol{g}}({\boldsymbol{k}})}{\left|{\boldsymbol{g}}({\boldsymbol{k}})\right|}\left[\Delta_{\uparrow\downarrow}(k)-\Delta_{\downarrow\uparrow}(k)\right], 
\label{eq_symmetry}
\end{align}
where ${\boldsymbol{d}}(k)$ is the gap function for spin-triplet pairing state 
with three components 
\begin{align}
d_x=&-\frac{1}{2}\left[\Delta_{\uparrow\uparrow}(k)-\Delta_{\downarrow\downarrow}(k)\right], \\
d_y=&\frac{1}{2{\mathrm{i}}}\left[\Delta_{\uparrow\uparrow}(k)+\Delta_{\downarrow\downarrow}(k)\right], \\
d_z=&\frac{1}{2}\left[\Delta_{\uparrow\downarrow}(k)+\Delta_{\downarrow\uparrow}(k)\right], 
\end{align}
due to degree of freedom of spin. 
This relation is derived under two assumptions. 
One is that there is no inter-band pairing between the Fermi surface split 
by the RSOC. 
The other is that intra-band pairings on the inner and outer Fermi surface 
have same pairing symmetry. 
In the present calculation within the RPA, 
the former assumption is broken, $i.e.$ there is finite inter-band 
pairing between the Fermi surface split by the RSOC. 
Therefore, the previous discussion (\ref{eq_symmetry}) is not applicable. 
Actually, the previous and present studies give different conclusions. 
According to the previous study, spin-triplet $f$-wave pairing symmetry 
should be admixed with spin-singlet $d_{x^2-y^2}$-wave one. 
On the other hand, in the present study on the basis of the RPA, 
it is derived that spin-triplet ($S_z=\pm1$) $p$-wave pairing 
symmetry is admixed with spin-singlet $d_{x^2-y^2}$-wave one 
in the region where $U$ ($V$) is large (small) 
as shown in Fig. \ref{fig_U-V-phase_lambda03}. 
Thus, in the case where the above two assumptions do not hold, 
we must take into consideration the dispersion relation 
and the pairing interaction. 
%
%
\section{SUMMARY}
\label{sec_summary}
In order to study the pairing symmetry in non-centrosymmetric superconductors, 
we have solved the linearized ${\acute{\mathrm{E}}}$liashberg's equation 
on the two-dimensional extended Hubbard model in the presence of the RSOC 
within the RPA. 
We found that three types of the pairing symmetries appeared in the 
$U$-$V$ phase diagram in the presence of the RSOC. 
In the region where $U$ ($V$) is large (small), 
$p+D_{x^2-y^2}$-wave pairing symmetry, which is predominantly 
spin-singlet $d_{x^2-y^2}$-wave one admixed with spin-triplet ($S_z=\pm1$) 
$p$-wave one, is the most stable. 
In the region where $U$ and $V$ are intermediate, 
$d_{x^2-y^2}+F$-wave pairing symmetry, which is predominantly 
spin-triplet ($S_z=\pm1$) $f$-wave one admixed with spin-singlet 
$d_{x^2-y^2}$-wave one, is the most stable. 
In the region where $U$ ($V$) is small (large), 
$p+D_{xy}$-wave pairing symmetry, which is predominantly 
spin-singlet $d_{xy}$-wave one admixed with spin-triplet ($S_z=\pm1$) 
$p$-wave one, is the most stable. 
\par
From analytical study, we found that pairing symmetry of an admixed 
subcomponent of the gap function depends on not only that of 
the predominant component but also the dispersion relation and 
momentum/space dependence of the pairing interaction. 
Amplitude of the admixed subcomponent of the gap function depends 
on the dispersion relation, $i.e.$ the hopping $t$ and the chemical 
potential $\mu$, as follows. 
For the $p+D_{x^2-y^2}$-wave pairing symmetry, amplitude of 
the admixed spin-triplet ($S_z=\pm1$) $p$-wave subcomponent increases 
with $\left| t\right|$. 
For the $d_{x^2-y^2}+F$-wave pairing symmetry, amplitude of 
the admixed spin-singlet $d_{x^2-y^2}$-wave subcomponent increases 
with $\left|\mu\right|$. 
For the $p+D_{xy}$-wave pairing symmetry, amplitude of 
the admixed spin-triplet ($S_z=\pm1$) $p$-wave subcomponent increases 
with $\left|\mu\right|$. 
%
%
\section{ACKNOWLEDGMENTS}
This work is supported by Grant-in-Aid for Young Scientists (B) No. 22740222 
and the "Topological Quantum Phenomena" (No. 22103005) 
Grant-in-Aid for Scientific Research on Innovative Areas from the Ministry 
of Education, Culture, Sports, Science and Technology (MEXT) of Japan. 
One of the authors (K.S) has been supported by Research Fellowships of 
the Japan Society for the Promotion of Science for Young Scientists. 
%
%

\end{document}